\documentclass{jfm}

\usepackage[none]{hyphenat}

\usepackage[hidelinks]{hyperref}
\usepackage[utf8]{inputenc}
\usepackage{wrapfig}

\usepackage{xr}
\usepackage{graphicx}
\usepackage{epstopdf, epsfig}

\usepackage{mathchars}
\usepackage{amsmath,amsfonts,amssymb}

\usepackage{float}

\usepackage{caption}
\usepackage{subcaption}

\usepackage[sort&compress]{natbib}

\usepackage{multicol}
\usepackage{multirow}

\usepackage{setspace}

\newcommand{\LS}{\left[}
\newcommand{\RS}{\right]}
\newcommand{\LB}{\left(}
\newcommand{\RB}{\right)}

\usepackage{tikz}

\shorttitle{Stability of Taylor bubble motion}
\shortauthor{H. A. Abubakar and O. K. Matar}

\title{Taylor bubble motion in stagnant and flowing liquids in vertical pipes.  
Part II: Linear stability analysis}

\author{H. A. Abubakar\aff{1,2}
 \and O. K. Matar\aff{1}
\corresp{\email{o.matar@imperial.ac.uk}} 
 }

\affiliation{\aff{1}Department of Chemical Engineering, Imperial College London, London SW7 2AZ, UK
\aff{2}Department of Chemical Engineering, Ahmadu Bello University, Zaria 810107, Nigeria}

\begin{document}
\setlength{\abovedisplayskip}{10pt}
\setlength{\belowdisplayskip}{10pt}
\setlength{\abovedisplayshortskip}{0pt}
\setlength{\belowdisplayshortskip}{0pt}

\maketitle
\begin{abstract}
In this study, we examine the linear stability of an axisymmetric Taylor bubble moving steadily in a flowing liquid enclosed in a circular tube. 
Linearisation is performed about axisymmetric base states obtained in Part I of this study by \cite{Abubakar_Matar_2021_s}.
The stability is characterised by the dimensionless inverse viscosity $\left( Nf \right)$, E\"{o}tv\"{o}s  $\left( Eo \right)$, and Froude numbers $\left( U_m \right)$, the latter being based on the centreline liquid velocity. 
The analysis shows that there exist regions of $(Nf,Eo,U_m)$ space within which the bubble is unstable and assumes an asymmetric shape.   To elucidate the mechanisms underlying the instability, an energy budget analysis is carried out which reveals that perturbation growth is driven by the bubble pressure for $Eo \geq 100$, and by the tangential interfacial stress for $Eo < 100$. Examples of the asymmetric bubble shapes and their associated flow fields are also provided near the onset of instability for a wide range of $Nf$, $Eo$, and $U_m$.
 
\end{abstract}


\section{Introduction} \label{sec:introduction}
Slug formation in gas-liquid flows is characterised by intermittent flow of large gas bubbles, separated by liquid masses. In vertical circular tubes, these large  bubbles are bullet-shaped with diameter approximately equal to that of the tube, and are known as \textit{Taylor bubbles}. Because of the pseudo-periodic nature of the motion of these bubbles, the study of the behaviour of a single Taylor bubble is considered as a paradigm for  understanding the slug flow regime in vertical tubes. For this reason, extensive experimental \citep{Griffith_Wallis_1961,Moissis_Griffith_1962,White_Beardmore_1962,Nicklin_etal_1962,Campos_Carvalho_1988,Bugg_Saad_2002,Nogueira_etal_2006,
Llewellin_etal_2012,Rana_etal_2015,Pringle_etal_2015,Fershtman_etal_2017}, and a number of theoretical studies \citep{Dumitrescu_1943,Brown_1965,Collins_etal_1978,Funada_etal_2005,Fabre_2016} have been carried out to examine the steady bubble shape, rise speed, and velocity field surrounding the bubble. 
In addition, numerical simulations have also been performed a large proportion of which are for  axisymmetric bubbles \citep{Mao_Dukler_1990,Mao_Dukler_1991,Bugg_Saad_2002,Taha_cui_2006,Lu_Prosperetti_2009,Kang_2010}, with comparatively  fewer studies focusing on the fully three-dimensional case \citep{Lizarraga-Garcia_etal_2017,Anjos_etal_2014,Taha_cui_2006}. 

\begin{figure}
    \centering
    \begin{subfigure}[H]{0.16\linewidth}
        \includegraphics[width=\linewidth]{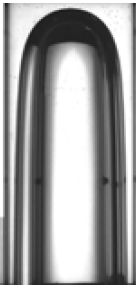}
        \caption{ }
        \label{fig:sym_bubble_fabre}
    \end{subfigure}
    \qquad \qquad \qquad 
    \begin{subfigure}[H]{0.175\linewidth}
        \includegraphics[width=\linewidth]{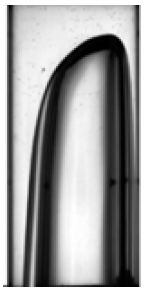}
        \caption{}
        \label{fig:asym_bubble_fabre}
    \end{subfigure}
    \caption{Taylor bubble shapes in a stagnant, (a), 
    and downward-flowing liquid, (b), 
    reproduced from \cite{Fabre_Figueroa-Espinoza_2014} } \label{fig:down_liq_stab}
\end{figure}

The shape of the Taylor bubble `nose' plays an important role in determining the rise speed, an important feature for developing predictive models for the slug flow regime \citep{Mao_Dukler_1990}. A Taylor bubble rising in a stagnant, or upward-flowing liquid, is generally axisymmetric and moves at constant speed in an inertia-dominated regime. This is not the case for Taylor bubbles moving in downward-flowing liquids, however, as experiments have confirmed the existence of a critical liquid velocity beyond which the bubble shape loses axisymmetry 
\citep{Fershtman_etal_2017,Fabre_Figueroa-Espinoza_2014,Polonsky_etal_1999,Martin_1976,Nicklin_etal_1962,Griffith_Wallis_1961}.  
An example of asymmetric bubble shapes in downward-flowing liquids is shown in Figure \ref{fig:down_liq_stab} in which it is seen that the bubble nose becomes distorted, and in an attempt to avoid the fast-moving fluid at the centre of the tube, the bubble moves closer to the tube wall \citep{Nicklin_etal_1962}, rising faster than it would have done had it remained axisymmetric 
\citep{Polonsky_etal_1999,Martin_1976}.
It was also noted by \cite{Martin_1976} that for a downward-flowing liquid, a stable axisymmetric Taylor bubble  can only be observed in tubes where surface tension effects are dominant, which is typical of  small diameter tubes characterised by low E\"{o}tv\"{o}s numbers; furthermore, the absolute value of the downward liquid velocity at which a bubble loses its axisymmetry decreases with increasing tube diameter.

Motivated by the aforementioned observations, \cite{Lu_Prosperetti_2006} carried out a linear stability analysis of a Taylor bubble moving in a downward-flowing liquid using potential flow theory and with negligible surface tension.  They demonstrated that an axisymmetric Taylor bubble rising in a liquid with a laminar velocity profile subjected to irrotational perturbations becomes unstable beyond a critical negative liquid velocity. 
Following on from the theoretical investigation of \cite{Lu_Prosperetti_2006} and the numerical simulation of \cite{Figueroa-Espinoza_Fabre_2011},  \cite{Fabre_Figueroa-Espinoza_2014} performed experimental investigations in tubes of diameters 20, 40, and 80 mm, complemented by numerical simulations using the boundary element method of \cite{Ha-Ngoc_Fabre_2006}. They showed that the radius of curvature of the bubble nose plays a key role in the stability of the Taylor bubble and that the onset of instability is dependent on the E\"{o}tv\"{o}s number. 

In this paper, we examine the stability of an axisymmetric steadily moving Taylor bubble in stagnant and flowing liquids, with particular attention given to the transition of the bubble shape from symmetric to asymmetric. To the best of our knowledge, the study of  \cite{Lu_Prosperetti_2006} represents the first attempt in the literature to understand the mechanism governing this transition using linear theory. However, experimental studies \citep{Fabre_Figueroa-Espinoza_2014} have shown that the onset of the instability is dependent on surface tension and by neglecting it, \cite{Lu_Prosperetti_2006} overestimate this onset. Moreover, existing studies of this transition have focused on the inertia-dominated regime, necessitating the need for  investigating the parameter spaces where both viscous and surface tension effects are important. Thus, we carry out a linear stability analysis to understand how the forces acting on the bubble, characterised by the dimensionless inverse viscosity, E\"{o}tv\"{o}s, and Froude numbers (the latter being based on the centreline liquid velocity) affect the loss of bubble axisymmetry. In addition, an energy budget analysis is carried out to determine the dominant, perturbation energy-producing terms that drive the instability. In a companion paper to the present work by \cite{Abubakar_Matar_2021_s}, we computed the steady state solutions for an axisymmetric Taylor bubble rising steadily in stagnant and flowing liquids for different values of the aforementioned dimensionless parameters. These solutions serve as the base states for the linear stability analysis carried out herein. We relate the conclusions drawn from the steady state analysis by \cite{Abubakar_Matar_2021_s} to the mechanisms governing the instability. 

The rest of this paper is organised as follows. In Section \ref{sec:formulation}, we provide details of the governing equations, weak formulation, normal modes analysis, and validation of the numerical technique used to carry out the computations. A discussion of the results of the linear stability, and of the energy budget analyses is provided in Sections 3 and 4, respectively. Finally, in Section 5, concluding remarks and perspectives for future research on remaining open questions related to Taylor bubble motion are provided.

\section{Problem formulation}
\label{sec:formulation}
\subsection{Governing equations and boundary conditions} 
\label{sec:linear_stability_governing_eqns}
\begin{figure}
\centering
	\includegraphics[width = 0.5\linewidth]{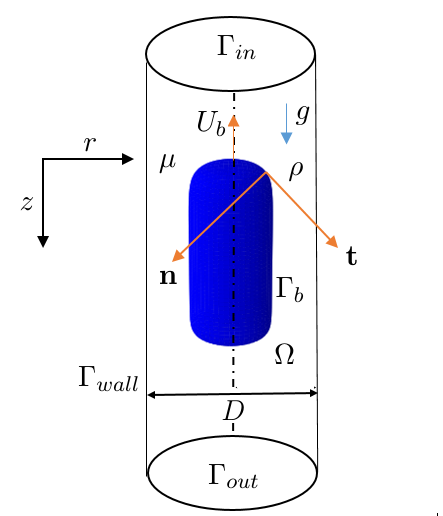}
	\caption{\label{fig:taylor_bubble_3D} Three-dimensional axisymmetric Taylor bubble moving with a steady speed $U_b$ through a liquid which is either stagnant or flowing (upwards or downwards) in a vertically-aligned tube of diameter $D$.}
\end{figure}

We consider a situation in which the two-dimensional axisymmetric steady state solution is revolved in the azimuthal direction to form a three-dimensional Taylor bubble, as shown in Figure \ref{fig:taylor_bubble_3D}. Here, we have adopted a cylindrical polar coordinate system $(r,\theta,z)$ in which $r$, $\theta$, and $z$ denote the radial, azimuthal, and axial coordinates, respectively. In the steady state analysis of \cite{Abubakar_Matar_2021_s}, the density, $\rho_g$, and viscosity, $\mu_g$, of the gas phase are assumed to be very small compared to their liquid counterparts, $\rho$ and $\mu$, respectively. Hence, the dynamics in the gas phase is approximated by a constant pressure ${P_b} $, its influence being restricted to the interface separating the phases designated by $\Gamma_b$, and, consequently, only the liquid flow field and the $P_b$ need to be determined. 

The flow is governed by the Navier-Stokes and continuity equations which are non-dimensionalised by the characteristic length, velocity, and pressure scales, $ D, \sqrt{gD}, \; \mbox{and} \;  \rho g D,  $ respectively, where $D$ is the tube diameter, and $g$ is the gravitational accleration. These equations are cast in a frame-of-reference translating with the velocity of the steadily-rising bubble nose $\mathbf{u}_b = -U_b \mathbf{i}_z $, in which $U_b$ is the constant rise speed, and given by 
 \begin{equation}
\frac{\partial \mathbf{u}}{\partial t} 
+ 
\left(\mathbf{u} \cdot \nabla \right)\mathbf{u}
-
\nabla \cdot \mathbf{T} 
= \mathbf{0},	\qquad  \text{in}  \qquad \Omega \LB t \RB \label{eq:momentum}
 \end{equation}
 \begin{equation}
 \nabla \cdot \mathbf{u} = 0,  \qquad  \text{in}  \qquad \Omega \LB t \RB \label{eq:continuity} 
 \end{equation}
where $\Omega$ denotes the domain of interest, $\mathbf{u}=(u_r, u_\theta, {u_z})$ where $u_r$, $u_\theta$, and $u_z$ are the components of the liquid velocity vector in the moving frame-of-reference $\mathbf{u}$ in the $r$, $\theta$, and $z$ directions, respectively, and $t$ denotes time;  $\mathbf{T}$ is the total stress tensor given by
\begin{equation}
\mathbf{T} = - {p} \mathbf{I} + 2   {Nf}^{-1} \mathbf{E}(\mathbf{u}), \label{eq:stress_tensor}
\end{equation}
in which $p$ represents the dynamic pressure,  $\mathbf{E}=(\nabla \mathbf{u}+\nabla\mathbf{u}^{T})/2$ is the rate of deformation tensor, and $\nabla=\mathbf{i}_r \frac{\partial}{\partial r}  + \mathbf{i}_\theta \frac{1}{r} \frac{\partial}{\partial \theta}  + \mathbf{i}_z \frac{\partial}{\partial z} $ is the gradient operator in cylindrical polar coordinates where $\mathbf{i}_r$, $\mathbf{i}_\theta$, and ${\mathbf{i}_z}$ are the unit vectors in the radial, azimuthal, and axial directions, respectively, and $\mathbf{I} $ is the unit tensor. In equation $\mathbf{T}$, the inverse viscosity number $Nf$ is defined as follows
\begin{equation}
    Nf = \frac{\rho \LB g D^3\RB ^{\frac{1}{2}}}{\mu}. \label{eq:inverse_vis_numb}
\end{equation}

The boundary of the domain, $\Gamma$, is divided into $\Gamma_{in}, \; \Gamma_{out}, \; \Gamma_{wall}$, and $\Gamma_b$ as shown in Figure  \ref{fig:taylor_bubble_3D}, with the subscripts `in', `out', `wall', and `b' representing  the inlet, outlet, wall, and bubble boundaries, respectively. At the wall, no-slip and no-penetration boundary conditions are imposed:
%
\begin{equation}
\mathbf{u} = -\mathbf{u}_b \quad \mathrm{on \quad   \Gamma_{wall} }. \label{eq:wall_bc}
\end{equation} 
At the inlet, prescribed values, $\mathbf{u}_{in}$, are specified for the velocity components: 
%
\begin{equation}
\mathbf{u} = \mathbf{u}_{in} - \mathbf{u}_b \quad \mathrm{on \quad   \Gamma_{in} }.  \label{eq:inlet_bc}
\end{equation} 
At the domain outlet, we impose the following conditions:
\begin{subequations}
\begin{align}
\mathrm{ \mathbf{n}\cdot\mathbf{T} \cdot \mathbf{n} } = 0, \label{eq:zero_normal_stress_bc} \\
\mathrm{ \LB \mathbf{I} - \mathbf{n} \otimes \mathbf{n} \RB } \cdot  \mathbf{u} = \mathbf{0}, \label{eq:zero_tangential_vel_bc}
\end{align}  
\end{subequations}
where $\mathbf{n}$ is the unit normal vector to the boundary. 
Finally, at the interface $\Gamma_b$, we set 
\begin{align}
\mathbf{n}\cdot\mathbf{T}\cdot{\mathbf{n}} + P_b - z  - {Eo}^{-1} \kappa = 0,  \label{eq:normal_stress_bc} \\
\mathbf{n}\cdot\mathbf{T} \times \mathbf{n}   = \mathbf{0},  \label{eq:tangential_stress_bc} \\
\frac{d\mathbf{r}_b}{dt} \cdot \mathbf{n} -\mathbf{u} \cdot \mathbf{n}  =0,   \label{eq:kinematic_bc}
\end{align}
where $ \kappa $ is the curvature of the interface,  $\mathbf{r}_b(t)$ represents the position vector of all the points on the portion of the boundary that corresponds to the interface $\Gamma_b $, and ${Eo}$ is the dimensionless E\"{o}tv\"{o}s  number expressed by 
\begin{equation}
    Eo = \frac{\rho g D^2}{\gamma}, \label{eq:eotvosh_numb}
\end{equation}
in which $\gamma$ denotes the (constant) surface tension. 
Equations  $\LB \ref{eq:normal_stress_bc} \RB $-$\LB \ref{eq:kinematic_bc} \RB$ correspond to the normal stress, tangential stress, and kinematic boundary conditions, respectively. Note that gravity appears in $\LB \ref{eq:normal_stress_bc} \RB $ as $z$ because the hydrostatic component of the pressure has been subtracted from the total pressure, leaving only the hydrodynamic part.

\subsection{Weak forms and perturbation equations}
\label{sec:linear_stability_model}
We begin the model development from the weak forms of the momentum and continuity equations (see appendix \ref{sec:weak_formulation}):
\begin{align}
%
&\int_{V}
\left\{
\frac{\partial {\mathbf{u}}}{\partial t} \cdot \mathbf{\Phi} 
+
\left[ 
\left({\mathbf{u}} \cdot \nabla \right)\mathbf{u}
\right] 
\cdot \mathbf{\Phi}
+
2 {Nf}^{-1} \mathbf{E}\left({\mathbf{u}} \right) 
: \mathbf{E}\left(\mathbf{\Phi} \right)  
-
{p} \left( \nabla \ldotp \mathbf{\Phi} \right)
\right\}dV,
  \nonumber \\
& \quad {} -
\int_{A_b}
\left\{ 
\left[ 
{Eo}^{-1}\kappa  + \mathrm{z}  - \mathrm{P}_{\mathrm{b}}
\right] \mathbf{n} \cdot \mathbf{\Phi}
\right\}dA_b
=0,  \label{eq:ls_weak_form_momentum}
\end{align}
\begin{equation}
\int_{V}
\left\{
 \left( \nabla \ldotp \mathbf{u} \right)\varphi 
 \right\}dV = 0.  \label{eq:ls_weak_form_continuity} 
 \end{equation}
Also, we have the weak form of the kinematic boundary condition:  
 \begin{equation}
 \int_{A_b}
\left\{ 
\LS
{\frac{d\mathbf{r}_b}{dt} \cdot \mathbf{n} -\mathbf{u} \cdot \mathbf{n} }
\RS \xi  
\right\} dA_b
=0.   \label{eq:ls_weak_form_kinematic_bc}
\end{equation}
In equations \eqref{eq:ls_weak_form_momentum}-\eqref{eq:ls_weak_form_kinematic_bc}, $\mathbf{\Phi} $, $\varphi$, and $\xi$ denote the test functions for the velocity vector, pressure, and interface deformation magnitude, respectively.

Similar to domain perturbation, we assume that the three-dimensional base flow domain is perturbed by the addition of infinitesimal deformation field $\tilde{\mathbf{x}}$ to its position vector and that equations \eqref{eq:ls_weak_form_momentum}-\eqref{eq:ls_weak_form_kinematic_bc} are valid on the three-dimensional perturbed domain. A similar approach has been adopted in the three-dimensional linear stability analysis of coating flow problems by \cite{carvalho_scriven_1999,Christodoulou_scriven_1988}. The perturbed three-dimensional domain can then be linearised around the base state three-dimensional axisymmetric domain, the deformation field restricted to the interface, just as it is expected in the classical linear stability approach; finally, the linearisation of the perturbed flow field variables about the base state solution is carried out. This approach to the derivation of the linear stability model can be seen as an extension of the total linearisation method used in solving two-dimensional viscous free boundary problems \citep{Kruyt_etal_1988,Cuvelier_Schulkes_1990} to three dimensions. 

Let the position vector of the perturbed domain be written as 
\begin{equation}
\mathbf{r} = \mathbf{r}^0 + \epsilon \tilde{\mathbf{x}}, \label{eq:ls_interface_perturbation}
\end{equation}
where $\mathbf{r}^0 = (r^0,\theta^0,z^0)$  represents the position vector of the unperturbed three-dimensional base flow domain, $\tilde{\mathbf{x}} = (\tilde{x}_r,\tilde{x}_\theta,\tilde{x}_z) $ is a deformation field defined over the entire base flow domain, and $\epsilon \ll 1$ to signify the infinitesimally small nature of the applied perturbations.  The linearised elemental volume of the perturbed three-dimensional domain is given as \citep{Cairncross_etal_2000, carvalho_scriven_1999}
\begin{eqnarray}
    dV=rdr d\theta dz &=& (1+\nabla\cdot \tilde{\mathbf{x}})r^0dr^0 d \theta^0 dz^0\nonumber\\
                      &=& (1+\nabla\cdot \tilde{\mathbf{x}})d\Omega^0 d\theta^0;
                      \label{eq:ls_elemental_volume_linearised}
\end{eqnarray}
thus, we can relate the elemental volume in the perturbed three-dimensional domain to the base flow two-dimensional axisymmetric domain, $d \Omega^0 = r^0dr^0dz^0$.
Similarly, an elemental area on the perturbed interface in the three-dimensional domain, $dA_b$,  can be related to base flow length of arc, $\Gamma^0_b$ in the two-dimensional axisymmetric domain:
\begin{equation}
d A_b = \LB 1 + \nabla_s \cdot \tilde{\mathbf{x}}\RB d\Gamma^0_b d\theta^0,	\label{eq:ls_elemental_area_linearised}
\end{equation} 
where $\nabla_s$ is the surface gradient operator; the interface terms can be linearised as follows
\begin{subequations}	\label{eq:ls_surface_terms_linearised}
\begin{align}
\mathbf{n} &= \mathbf{n}^0 + \epsilon \tilde{\mathbf{n}}, \\
\kappa &= \kappa^0 + \epsilon \tilde{\kappa}, \\
\mathbf{\Phi} &=  \mathbf{\Phi} + \epsilon \left( \tilde{\mathbf{x}}\cdot \nabla \right){\mathbf{\Phi}}, \\
\mathbf{u} &=  \mathbf{u} + \epsilon \left( \tilde{\mathbf{x}}\cdot \nabla \right){\mathbf{u}};
\end{align}
\end{subequations}
here, $\mathbf{n}^0$ and $\kappa^0$ are the base state  interface normal vector and curvature, and $\tilde{\mathbf{n}}$ and $\tilde{\kappa}$  represent the normal vector and curvature perturbations, respectively 
(see Appendix \ref{sec:curvature_linearisation}):
\begin{equation}
\tilde{\mathbf{n}} = - \mathbf{t}^0 \LB \mathbf{n}^0 \cdot \frac{d \tilde{\mathbf{x}}}{d s^0}\RB - \frac{\mathbf{n}^0}{r^0} \cdot \frac{\partial \tilde{\mathbf{x}}}{\partial \theta^0}, \mathbf{i}_\theta  \label{eq:normal_linearised_simplified}
\end{equation}
\begin{multline}
\tilde{\kappa} = \frac{1}{r^0} \frac{d}{d s^0} \LS r^0 \LB \mathbf{n}^0 \cdot \frac{d \tilde{\mathbf{x}}}{d s^0}\RB \RS + 2 \LB \mathbf{t}^0 \cdot  \frac{d \tilde{\mathbf{x}}}{d s^0}  \RB \LB \mathbf{t}^0 \cdot \frac{d \mathbf{n}^0}{d s^0} \RB  \\
+  \frac{\mathbf{n}^0}{{r^0}^2} \cdot \frac{\partial^2 \tilde{\mathbf{x}}}{\partial {\theta^0}^2} + \frac{\tilde{x}_r n_r^0}{{r^0}^2} - \frac{d \mathbf{n}^0}{d s^0} \cdot \frac{d \tilde{\mathbf{x}}}{d s^0},  \label{eq:curvature_perturb0}
\end{multline}
where $\frac{d }{d s^0} = \mathbf{t} \cdot \nabla$ is the derivative along the arc length $s$ on the base state interface.
Substitution into equations \eqref{eq:ls_weak_form_momentum}-\eqref{eq:ls_weak_form_kinematic_bc} of equations \eqref{eq:ls_interface_perturbation}-\eqref{eq:ls_surface_terms_linearised}, together with the flow field perturbations
\begin{subequations}
\begin{align}
\mathbf{u} = \mathbf{u}^0 + \epsilon \tilde{\mathbf{u}}, \\
{p} = {p}^0 + \epsilon \tilde{{p}},
\end{align}
\end{subequations}
followed by neglecting all terms of order $\epsilon^2$ respectively yields the following leading order momentum, continuity, and kinematic condition equations 
%
%
\begin{align}
&\int_0^{2 \pi}
\left \{
\int_{\Omega^0}
\left\{
\left[ 
\left({\mathbf{u}^0} \cdot \nabla \right)\mathbf{u}^0
\right] 
\cdot \mathbf{\Phi}
+
2 {Nf}^{-1} \mathbf{E}\left({\mathbf{u}^0} \right) 
: \mathbf{E}\left(\mathbf{\Phi} \right)
-
{p}^0 \left( \nabla \ldotp \mathbf{\Phi} \right)
\right\}d\Omega^0 
\right. \nonumber \\
& \left.
\quad {}
-	
\int_{\Gamma_b^0}
\left\{ 
\left[ 
{Eo}^{-1} \kappa^0  + z^0  - P_b^0
\right] \mathbf{n}^0 \cdot \mathbf{\Phi}
\right\}d\Gamma_b^0
\right \} d \theta^0,
\label{eq:ls_weak_form_momentum_leading_order}
\end{align}
%
%
\begin{align}
 & \int_0^{2 \pi}
\int_{\Omega^0}
\left \{
\left\{
 \left( \nabla \ldotp \mathbf{u}^0 \right)  \varphi  
\right \} 
\LS
1 + \nabla \cdot \tilde{\mathbf{x}}
\RS
 \right\}d\Omega^0  
 d \theta^0,
 \label{eq:ls_weak_form_continuity_leading_order}
\end{align}
%
%
 \begin{align}
& \int_0^{2 \pi}
 \int_{\Gamma_b^0}
\left\{
\left\{ 
\LS
{\frac{d\mathbf{r}_b^0}{dt} \cdot \mathbf{n}^0 -\mathbf{u}^0 \cdot \mathbf{n}^0 }
\RS \xi  
\right\}
\LS
1 + \nabla_s \cdot \tilde{\mathbf{x}}
\RS
\right \} 
d\Gamma_b^0  d \theta^0.
\label{eq:ls_weak_form_kinematic_bc_leading_order}
\end{align}
It is also possible to write the following equations at $O(\epsilon)$ to yield equations that feature the perturbation variables: 
the momentum conservation equation,
%
%
\begin{align}
&
\int_0^{2\pi}
\int_{\Omega^0}
\left\{
\frac{\partial \tilde{\mathbf{u}}}{\partial t} \cdot \mathbf{\Phi} 
+
{
\left[ 
\left({\mathbf{u}^0} \cdot \nabla \right) \tilde{\mathbf{u}} + \left(\tilde{\mathbf{u}} \cdot \nabla \right)\mathbf{u}^0 
\right] 
\cdot \mathbf{\Phi} 
}
+
2 {Nf}^{-1} \mathbf{E}\left(\tilde{\mathbf{u}} \right) 
: \mathbf{E}\left(\mathbf{\Phi} \right)
\right\}
d\Omega^0 d \theta^0  \nonumber \\
& \quad {} 
-\int_0^{2\pi}
\int_{\Omega^0}
\left\{
\tilde{p} \left( \nabla \ldotp \mathbf{\Phi} \right)
\right\}
d\Omega^0 d \theta^0  \nonumber \\
& \quad {} 
- 
\int_0^{2\pi}
\int_{\Gamma_b^0}
\left\{ 
\left[ 
{Eo}^{-1} \tilde{\kappa}  + \tilde{z}
\right] \mathbf{n}^0 \cdot \mathbf{\Phi}
\right\}d\Gamma_b^0 d \theta^0 \nonumber \\
& \quad {}
+
{ 
\int_0^{2\pi}
\int_{\Gamma^0_b}
\tilde{\mathbf{x}} \cdot \mathbf{n}^0
\left\{
{
\left[ 
\left({\mathbf{u}^0} \cdot \nabla \right) \mathbf{u}^0  
\right] 
\cdot \mathbf{\Phi} 
}
+
2 {Nf}^{-1} \mathbf{E}\left(\mathbf{u}^0 \right) 
: \mathbf{E}\left(\mathbf{\Phi} \right)
-
p^0 \left( \nabla \ldotp \mathbf{\Phi} \right)
\right\}
d\Gamma^0_b d \theta^0
} \nonumber \\
& \quad {}
-
\int_0^{2\pi}
\int_{\Gamma_b^0}
\left\{ 
\left[ 
{Eo}^{-1} \kappa^0  + z^0 - P_b^0
\right] \LS \tilde{\mathbf{n}} \cdot \mathbf{\Phi} +  \LS \left( \tilde{\mathbf{x}}\cdot \nabla \right){\mathbf{\Phi}} \RS \cdot \mathbf{n}^0 + \LB \nabla_s \cdot \tilde{\mathbf{x}} \RB \mathbf{n}^0 \cdot \mathbf{\Phi} \RS
\right\}d\Gamma_b^0 d \theta^0 \nonumber \\
& \quad {} = 0,		 \label{eq:ls_weak_form_momentum_linearised}
\end{align}	
where $\mathbf{u}^0$, $p^0$, and $P_b^0$ represent the base flow solutions for the variables; and $\tilde{\mathbf{u}}$ and $\tilde{p}$ denote the perturbations to the flow field variables, and the last two lines of \eqref{eq:ls_weak_form_momentum_linearised} are due to the linearisation of the domain and boundary terms of equation \eqref{eq:ls_weak_form_momentum} where we have used Gauss's divergence theorem to restrict the deformation to the interface as it is expected in classical linear stability formulation; 
the continuity equation, 
%
%
\begin{equation}
 \int_0^{2 \pi}
\int_{\Omega^0}
\left \{
 \left( \nabla \ldotp \tilde{\mathbf{u}} \right)  \varphi   
 \right\}d\Omega^0  
 d \theta^0  = 0,	\label{eq:ls_weak_form_continuity_linearised}
\end{equation}
and the kinematic condition: 
%
%
 \begin{equation}
\int_0^{2 \pi}
\int_{\Gamma_b^0}
\left\{ 
\LS
{\frac{d \tilde{\mathbf{x}}}{dt} \cdot \mathbf{n}^0 -\tilde{\mathbf{u}} \cdot \mathbf{n}^0 - \mathbf{u}^0 \cdot \tilde{\mathbf{n}}  
- 
\LS
\left( \tilde{\mathbf{x}}\cdot \nabla \right) \mathbf{u}^0
\RS \cdot \mathbf{n}^0
}
\RS \xi  
\right\} d\Gamma_b^0  d \theta^0 
= 0. \label{eq:ls_weak_form_kinematic_bc_linearised}
\end{equation}
Simplifying equations \eqref{eq:ls_weak_form_momentum_linearised}-\eqref{eq:ls_weak_form_kinematic_bc_linearised} further by substituting for $\tilde{\mathbf{n}}$ and $\tilde{\kappa}$ using equations \eqref{eq:normal_linearised_simplified} and \eqref{eq:curvature_perturb0}, respectively, and taking the deformation field  to be of the form 
$\tilde{\mathbf{x}} = \tilde{h} \mathbf{n}^0$,
since the deformation has  been restricted to the interface and making use of the relations 
\begin{subequations}
\begin{align}
\mathbf{u}^0  = \LB \mathbf{u}^0 \cdot \mathbf{n}^0 \RB \mathbf{n}^0 &+  \LB \mathbf{u}^0 \cdot \mathbf{t}^0 \RB \mathbf{t}^0, \\
{\nabla} = \LB \mathbf{I} - \mathbf{n}^0 \otimes \mathbf{n}^0 \RB \cdot \mathbf{\nabla} &+ \LB \mathbf{n}^0 \otimes \mathbf{n}^0 \RB \cdot {\nabla},
\end{align}
\end{subequations}
equations \eqref{eq:ls_weak_form_momentum_linearised}-\eqref{eq:ls_weak_form_kinematic_bc_linearised}, after some algebra, can be expressed as follows
%
%
\begin{align}
&
\int_0^{2\pi}
\int_{\Omega^0}
\left\{
\frac{\partial \tilde{\mathbf{u}}}{\partial t} \cdot \mathbf{\Phi} 
+
{
\left[ 
\left({\mathbf{u}^0} \cdot \nabla \right) \tilde{\mathbf{u}} + \left(\tilde{\mathbf{u}} \cdot \nabla \right)\mathbf{u}^0 
\right] 
\cdot \mathbf{\Phi} 
}
+
2 {Nf}^{-1} \mathbf{E}\left(\tilde{\mathbf{u}} \right) 
: \mathbf{E}\left(\mathbf{\Phi} \right)
\right\}
d\Omega^0 d \theta  \nonumber\\
& \quad {}
-\int_0^{2\pi}
\int_{\Omega^0}
\left\{
\tilde{p} \left( \nabla \ldotp \mathbf{\Phi} \right)
\right\}
d\Omega^0 d \theta \nonumber\\
& \quad {} -
\int_0^{2\pi}
\int_{\Gamma_b^0} 
{Eo}^{-1} 
\left \{
-\frac{d \tilde{h}}{ds} \left[ \mathbf{n} \cdot \frac{d \mathbf{\Phi}}{ds} - \kappa_a \left( \mathbf{t} \cdot \mathbf{\Phi} \right)\right]
+
 \left[ \tilde{h} \LB \kappa_a^2 + \kappa_b^2 \RB + \frac{1}{r^2} \frac{\partial^2 \tilde{h} }{\partial \theta^2} \right] \mathbf{n} \cdot \mathbf{\Phi}
\right \} 
d\Gamma_b^0 d \theta
\nonumber\\
& \quad {}
-
\int_0^{2\pi}
\int_{\Gamma_b^0}
\left\{ 
\tilde{h} n_z 
\LB
 \mathbf{n} \cdot \mathbf{\Phi}
 \RB
\right\}d\Gamma_b^0 d \theta \nonumber\\
& \quad {} + 
\int_0^{2\pi}
\int_{\Gamma^0_b}
\tilde{h}
\left\{
{
\LB
\mathbf{u}^0 \cdot \mathbf{t}
\RB
\left[ 
\frac{d \mathbf{u}^0}{d s} \cdot \mathbf{\Phi}
\right] 
}
+
\LS -p^0 + 2 Nf^{-1} \LB \mathbf{t} \cdot \frac{d \mathbf{u}^0}{d s} \RB  \RS \LB \mathbf{t} \cdot \frac{d \mathbf{\Phi}}{d s} \RB  \right.
\nonumber\\
& \quad {}
\left.
+
\LS -p^0 + 2 Nf^{-1} \frac{u_r^0}{r}  \RS \LB \frac{\Phi_r}{r} + \frac{1}{r} \frac{\partial \Phi_\theta}{\partial \theta}\RB 
\right\}
d\Gamma^0_b d \theta
\nonumber\\
& \quad {}
+
\int_0^{2\pi}
\int_{\Gamma_b^0}
\left\{ 
\left[ 
{Eo}^{-1} \kappa  + z - P_b^0
\right] 
\LS 
\LB \mathbf{t} \cdot \mathbf{\Phi} \RB \frac{d \tilde{h}}{ds} + \frac{\Phi_\theta}{r} \frac{\partial \tilde{h}}{\partial \theta}
+
\tilde{h} \kappa \LB \mathbf{n} \cdot \mathbf{\Phi}\RB
\RS
\right\}d\Gamma_b^0 d \theta
= 0,		   \label{eq:ls_weak_form_momentum_perturbation}
\end{align}	
%
%
\begin{equation}
 \int_0^{2 \pi}
\int_{\Omega^0}
\left \{
 \left( \nabla \ldotp \tilde{\mathbf{u}} \right)  \varphi  
 \right\}d\Omega  
 d \theta  = 0,	 \label{eq:ls_weak_form_continuity_perturbation}
 \end{equation}
%
%
 \begin{equation}
\int_0^{2 \pi}
\int_{\Gamma_b^0}
\left\{ 
\LS
{\frac{d \tilde{h}}{dt} -\tilde{\mathbf{u}} \cdot \mathbf{n} + \LB \mathbf{u}^0 \cdot \mathbf{t} \RB   \frac{d \tilde{h}}{ds} 
- 
\tilde{h} \LB \frac{d \mathbf{u}^0}{dn} \cdot \mathbf{n} \RB
}
\RS \xi  
\right\} d\Gamma_b^0  d \theta  = 0, 	 \label{eq:ls_weak_form_kinematic_bc_perturbation}
 \end{equation}
where $\frac{d}{dn} = \LB \mathbf{n} \cdot \nabla \RB $ is the derivative in the normal direction.  
In equations \eqref{eq:ls_weak_form_momentum_perturbation}-\eqref{eq:ls_weak_form_kinematic_bc_perturbation}, we have suppressed the use of the superscript $`0'$ to designate base state quantities for the unit tangent and normal vectors for the sake of brevity. 

\subsection{Normal modes}
Let us take the following normal mode forms for the perturbation variables: 
\begin{subequations} \label{eq:ls_nomal_mode_perturbation_forms}
\begin{align}
\tilde{\mathbf{u}}\left( r,\theta,z, t \right) &=  \hat{\mathbf{u}} \LB r,z \RB e^{ \LB \mathfrak{i} m \theta + \beta t \RB},  \label{eq:ls_nomal_mode_perturbation_forms_u}\\
\tilde{p}\left( r,\theta,z, t \right) &= \hat{p} \LB r,z \RB e^{ \LB \mathfrak{i} m \theta + \beta t \RB}, \label{eq:ls_nomal_mode_perturbation_forms_p} \\
\tilde{h}\LB s, \theta, t \RB &= \hat{h} \LB s \RB e^{ \LB \mathfrak{i} m \theta + \beta t \RB},   \label{eq:ls_nomal_mode_perturbation_forms_h}
\end{align}
\end{subequations}
and their corresponding test functions as
\begin{subequations}	\label{eq:ls_nomal_mode_test_function_forms}
\begin{align}
\mathbf{\Phi} \left( r, \theta, z\right) &= \bar{\mathbf{\Phi}} \left( r, z\right)e^{ \LB -\mathfrak{i} m \theta \RB}, \\
{\varphi} \left( r, \theta, z\right) &= \bar{\varphi} \left( r, z\right)e^{ \LB -\mathfrak{i} m \theta  \RB}, \\
\xi \left( s, \theta \right) &=  \bar{\xi}\left( s\right)e^{ \LB -\mathfrak{i} m \theta \RB},
\end{align}
\end{subequations}
where $\hat{\mathbf{u}}$, $\hat{p}$, and $\hat{h}$ are complex functions of space representing the amplitude of the velocity, pressure, and interface deformation perturbations, respectively; $m$ is a dimensionless (integer) wave number in the azimuthal direction $\theta$; 
$\beta=\beta_R + i \beta_I$ is the complex growth rate which can be decomposed into its real $\beta_R$ and imaginary $\beta_I$ parts denoting the temporal growth rate and frequency, respectively: if $\beta_R$ is positive (negative), the disturbance grows (decays) exponentially in time and the base flow is linearly unstable (stable); if $\beta_R$ is zero, the disturbance is neutrally stable. Substituting equations \eqref{eq:ls_nomal_mode_perturbation_forms} and \eqref{eq:ls_nomal_mode_test_function_forms} into \eqref{eq:ls_weak_form_momentum_perturbation}-\eqref{eq:ls_weak_form_kinematic_bc_perturbation}, separating the momentum equation into its components, yields the following equations governing the normal mode evolution of the perturbations as a function of $Nf$, $Eo$, $U_{m}$, and $m$: 
\begin{align}	 
& \int_{\Omega^0}
\left\{
\beta \hat{u}_r {\bar{\Phi}_r} 
+
\left[
\hat{u}_r \frac{\partial {u}_r^0}{\partial r} + \hat{u}_z \frac{\partial {u}_r^0}{\partial z}
+
{u}_r^0\frac{\partial\hat{u}_r}{\partial r} + {u}_z^0 \frac{\partial \hat{u}_r}{\partial z}
\right] {\bar{\Phi}_r} 
+ 
Nf^{-1} 
\LS
2 \frac{\partial \hat{u}_r}{\partial r} \frac{\partial \bar{\Phi}_r}{\partial r}
\right. \right. 
\nonumber\\
& \left. \left.
\quad {} 
+ \LB 2 + m^2 \RB \frac{\hat{u}_r \bar{\Phi}_r}{r^2} 
+ 3 \mathfrak{i} m \frac{\hat{u}_\theta \bar{\Phi}_r}{r^2} 
- \mathfrak{i} m \frac{\bar{\Phi}_r}{r} \frac{\partial \hat{u}_\theta}{\partial r}
+ \frac{\partial \bar{\Phi}_r}{\partial z} \LB \frac{\partial \hat{u}_r}{\partial z} + \frac{\partial \hat{u}_z}{\partial r} \RB
\RS
\right. 
\nonumber\\
& \left. 
\quad {} 
-
\hat{p} \LB \frac{\partial \bar{\Phi}_r}{\partial r} + \frac{{\bar{\Phi}_r}}{r} \RB
\right\}d\Omega^0 
\nonumber \\ 
& \quad {} 
-
\int_{\Gamma_b^0} 
{Eo}^{-1} 
\left \{
-\frac{d \hat{h}}{ds} \left[ n_r \frac{d \bar{\Phi}_r}{ds} - \kappa_a \left( t_r  \bar{\Phi}_r \right)\right]
+
\hat{h} \left[\kappa_a^2 + \kappa_b^2 - \frac{m^2}{r^2} \right] n_r  \bar{\Phi}_r
\right \} 
d\Gamma_b^0 
\nonumber \\ 
& \quad {} 
-
\int_{\Gamma_b^0}
\left\{ 
\hat{h} n_z 
\LB
 n_r  \bar{\Phi}_r
 \RB
\right\}d\Gamma_b^0 
\nonumber \\
& \quad {}  
+ 
\int_{\Gamma^0_b}
\hat{h}
\left\{
{
\LB
{u}^0_r {t}_r
\RB
\left[ 
\frac{d {u}^0_r}{d s}  \bar{\Phi}_r
\right] 
} 
+
\LS -p^0 + 2 Nf^{-1} \LB {t}_r \frac{d {u}^0_r}{d s} \RB  \RS \LB {t}_r  \frac{d \bar{\Phi}_r}{d s} \RB 
\right. 
\nonumber\\
& \left. 
\quad {} 
+
\LS -p^0 + 2 Nf^{-1} \frac{u_r^0}{r}  \RS \LB \frac{\bar{\Phi}_r}{r} \RB 
\right\}
d\Gamma^0_b  \nonumber
\\
& \quad {} 
+
\int_{\Gamma_b^0}
\left\{ 
\left[ 
{Eo}^{-1} \kappa  + z - P_b^0
\right] 
\LS 
\LB {t}_r  \bar{\Phi}_r \RB \frac{d \hat{h}}{ds} +
\hat{h} \kappa \LB {n}_r  \bar{\Phi}_r\RB
\RS
\right\}d\Gamma_b^0 
= 0,			\label{eq:ls_weak_form_normal_mode_perturbation_radial}
\end{align}
%
\begin{align}	
& \int_{\Omega^0}
\left\{
\beta \hat{u}_\theta {\bar{\Phi}_\theta} 
+
\LS
{u}_r^0 \frac{\partial \hat{u}_\theta}{\partial r}  + {u}_z^0 \frac{\partial \hat{u}_\theta}{\partial z} + \frac{{u}_r^0 \hat{u}_\theta}{r} 
\RS {\bar{\Phi}_\theta} 
+ 
Nf^{-1} 
\LS
\LB 1 + 2m^2 \RB \frac{\hat{u}_\theta \bar{\Phi}_\theta}{r^2} 
+
\frac{\partial \hat{u}_\theta}{\partial z} \frac{\partial \bar{\Phi}_\theta}{\partial z}
\right. \right. 
\nonumber \\  
& \quad {}  \left. \left.
+
\frac{\partial \hat{u}_\theta}{\partial r} \frac{\partial \bar{\Phi}_\theta}{\partial r} 
- 
\LB 
\frac{\hat{u}_\theta}{r} \frac{\partial \bar{\Phi}_\theta}{\partial r} + \frac{ \bar{\Phi}_\theta}{r} \frac{\partial \hat{u}_\theta}{\partial r} \RB
+ 
\mathfrak{i} m 
\LB 
\frac{\hat{u}_r}{r} \frac{\partial \bar{\Phi}_\theta}{\partial r} + \frac{\hat{u}_z}{r} \frac{\partial \bar{\Phi}_\theta}{\partial z} 
\RB
- 
3 \mathfrak{i} m \frac{\hat{u}_r\bar{\Phi}_\theta}{r^2} 
\RS
\right.  
\nonumber \\  
& \quad {} \left. 
-
p \LB -\mathfrak{i} m  \frac{\bar{\Phi}_\theta}{r} \RB
\right\}d\Omega^0  
+ 
\int_{\Gamma^0_b}
\hat{h}
\left\{
\LS -p^0 + 2 Nf^{-1} \frac{u_r^0}{r}  \RS \LB -\mathfrak{i} m  \frac{\bar{\Phi}_\theta}{r} \RB 
\right\}
d\Gamma^0_b
\nonumber \\
& \quad {}  
+
\int_{\Gamma_b^0}
\left\{ 
\hat{h}
\left[ 
{Eo}^{-1} \kappa  + z - P_b^0
\right] 
\LS 
\mathfrak{i} m  \frac{\bar{\Phi}_\theta}{r}
\RS
\right\}d\Gamma_b^0
= 0,			\label{eq:ls_weak_form_normal_mode_perturbation_azimuth}
\end{align}
%
\begin{align} 
& \int_{\Omega^0}
\left\{
\beta \hat{u}_z {\bar{\Phi}_z} 
+
\LS
\hat{u}_r \frac{\partial {u}_z^0}{\partial r} + \hat{u}_z \frac{\partial {u}_z^0}{\partial z}
+
{u}_r^0\frac{\partial\hat{u}_z}{\partial r} + {u}_z^0 \frac{\partial \hat{u}_z}{\partial z}
\RS {\bar{\Phi}_z} 
+ 
Nf^{-1} 
\LS
2 \frac{\partial \hat{u}_z}{\partial z} \frac{\partial \bar{\Phi}_z}{\partial z}
+  m^2 \frac{\hat{u}_z \bar{\Phi}_z}{r^2} 
\right. \right. 
\nonumber \\ 
& \quad {}   
\left. \left.
- \mathfrak{i} m \frac{\bar{\Phi}_z}{r} \frac{\partial \hat{u}_\theta}{\partial z}
+ \frac{\partial \bar{\Phi}_z}{\partial r} \LB \frac{\partial \hat{u}_r}{\partial z} + \frac{\partial \hat{u}_z}{\partial r} \RB
\RS
-
\hat{p} \LB \frac{\partial \bar{\Phi}_z}{\partial z}  \RB
\right\}d\Omega^0 
\nonumber \\
& \quad {}   
-
\int_{\Gamma_b^0} 
{Eo}^{-1} 
\left \{
-\frac{d \hat{h}}{ds} \left[ n_z \frac{d \bar{\Phi}_z}{ds} - \kappa_a \left( t_z  \bar{\Phi}_z \right)\right]
+
\hat{h} \left[\kappa_a^2 + \kappa_b^2 - \frac{m^2}{r^2} \right] n_z  \bar{\Phi}_z
\right \} 
d\Gamma_b^0
\nonumber \\
& \quad {}  
-
\int_{\Gamma_b^0}
\left\{ 
\hat{h} n_z 
\LB
 n_z  \bar{\Phi}_z
 \RB
\right\}d\Gamma_b^0 
\nonumber \\ 
& \quad {}  
+ 
\int_{\Gamma^0_b}
\hat{h}
\left\{
{
\LB
{u}^0_z {t}_z
\RB
\left[ 
\frac{d {u}^0_z}{d s}  \bar{\Phi}_z
\right] 
}
+
\LS -p^0 + 2 Nf^{-1} \LB {t}_z \frac{d {u}^0_z}{d s} \RB  \RS \LB {t}_z  \frac{d \bar{\Phi}_z}{d s} \RB 
\right\}
d\Gamma^0_b
\nonumber \\
& \quad {}   
+
\int_{\Gamma_b^0}
\left\{ 
\left[ 
{Eo}^{-1} \kappa  + z - P_b^0
\right] 
\LS 
\LB {t}_z  \bar{\Phi}_z \RB \frac{d \hat{h}}{ds} +
\hat{h} \kappa \LB {n}_z  \bar{\Phi}_z\RB
\RS
\right\}d\Gamma_b^0
= 0,				\label{eq:ls_weak_form_normal_mode_perturbation_axial}
\end{align}	
\begin{equation}	
\int_{\Omega^0}
\left\{
\LS
\frac{\partial \hat{u}_r}{\partial r} + \frac{\hat{u}_r}{r} - \mathfrak{i} m  \frac{\hat{u}_\theta}{r} + \frac{\partial \hat{u}_z}{\partial z}
\RS \bar{\varphi}
\right\}d\Omega^0
= 0,		\label{eq:ls_weak_form_normal_mode_perturbation_continuity}
\end{equation}
\begin{equation}
\int_{\Gamma_b^0}
\left\{ 
\LS
{
\beta \hat{h} -\hat{\mathbf{u}} \cdot \mathbf{n} + \LB \mathbf{u}^0 \cdot \mathbf{t} \RB   \frac{d \hat{h}}{ds} 
- 
\hat{h} \LB \frac{d \mathbf{u}^0}{dn} \cdot \mathbf{n} \RB
}
\RS \bar{\xi}  
\right\} d\Gamma_b^0 = 0. \label{eq:ls_weak_form_normal_mode_perturbation_kinematic_bc}
\end{equation}

The combined finite element forms for the perturbations equations   \eqref{eq:ls_weak_form_normal_mode_perturbation_radial}-\eqref{eq:ls_weak_form_normal_mode_perturbation_kinematic_bc} can be recast as a generalised eigenvalue problem 
\begin{equation}
\beta \mathbf{B} \mathbf{y}  = \mathbf{J}  \mathbf{y}, 	\label{eq:ls_algebraic_sys_eqns0}
\end{equation}
with $\beta$ being the eigenvalue, $\mathbf{B}$ the mass matrix,  $\mathbf{y}$ the eigenfunctions, and $\mathbf{J}$ the Jacobian matrix respectively given by
\begin{align}
{
\mathbf{B} 
= 
\begin{bmatrix}
{\mathbf{B}}_{r,r} &  0  &  0 & 0 & 0 \\
0 &   {\mathbf{B}}_{\theta,\theta}  &   0 & 0 &  0 \\
0 &   0 & {\mathbf{B}}_{z,z}  & 0 &  0  \\
0   & 0 &  0 & 0 &  0  \\
 0  & 0 &  0 & 0 &  {\mathbf{B}}_{h} 
\end{bmatrix}, 
}
\quad
{
\mathbf{y}
=
\begin{bmatrix}
\hat{\mathbf{v}}_r \\
\hat{\mathbf{v}}_\theta \\
\hat{\mathbf{v}}_z \\
\hat{\mathbf{p}} \\
\hat{\mathbf{h}}
\end{bmatrix},
} 
\quad 
\quad \nonumber \\
{
\mathbf{J} 
= 
-\begin{bmatrix}
\LB {\mathbf{C}}_{r,r} + {\mathbf{K}}_{r,r} \RB &   {\mathbf{K}}_{r,\theta}  &  \LB {\mathbf{C}}_{r,z} + {\mathbf{K}}_{r,z} \RB & -{\mathbf{Q}}_{r}^T &  {\mathbf{M}}_{r,h} \\
{\mathbf{K}}_{r,\theta}^T &   \LB {\mathbf{C}}_{\theta,\theta} + {\mathbf{K}}_{\theta,\theta} \RB &   {\mathbf{K}}_{z,\theta}^T & -{\mathbf{Q}}_{\theta}^T &  {\mathbf{M}}_{\theta,h} \\
\LB {\mathbf{C}}_{z,r} + {\mathbf{K}}_{r,z}^T \RB &    {\mathbf{K}}_{z,\theta} &  \LB {\mathbf{C}}_{z,z} + {\mathbf{K}}_{z,z} \RB & -{\mathbf{Q}}_{z}^T &  {\mathbf{M}}_{z,h}  \\
 -{\mathbf{Q}}_{r}   &  -{\mathbf{Q}}_{\theta} &  -{\mathbf{Q}}_{z} & 0 &  0  \\
  -{\mathbf{X}}_{r}  & 0 &  -{\mathbf{X}}_{z} & 0 &  {\mathbf{X}}_{h} 
\end{bmatrix}
} \nonumber	\label{eq:ls_algebraic_sys_eqns1}
\end{align}
where $\mathbf{B}_{i,i} \quad \forall \; i = r,\theta,z $ are  coefficient matrices for the $i$ component of velocity in the $i$ component of momentum equation. $\mathbf{C}_{i,j} \quad \text{and} \quad \mathbf{K}_{i,j}\quad \forall \; i = r,\theta,z $ are the convective and viscous coefficient matrices for the $j$ component of velocity in the $i$ component of momentum equation, respectively. $\mathbf{M}_{i,h} \quad \forall \; i = r,\theta,z $ are  coefficient matrices for interface deformation magnitude in the $i$ component of momentum equation. $\mathbf{Q}_{j} \quad \forall \; j = r,\theta,z $ are coefficient matrices for the $j$ component of velocity in the continuity equation, whose Hermitian transpose correspond to the coefficient matrices of pressure in the $j$ component of momentum equation. $\mathbf{X}_{j} \quad \forall \; j = r,z $ are coefficient matrices for the $j$ component of velocity in the kinematic boundary condition. $\mathbf{B}_{h} \quad \text{and} \quad \mathbf{X}_{h} $ are the coefficient matrices of interface deformation magnitude in the kinematic boundary condition for the mass and Jacobian matrices, respectively. The operator $\LB \cdot \RB^T$  denotes the Hermitian transpose. The full expressions for the coefficient matrices  in the mass and Jacobian matrices are given in Appendix \ref{sec:galerkin_method}. 

The boundary conditions at the inlet, wall, and  outlet reduce to the following conditions on the perturbations:
\begin{equation}
\hat{\mathbf{u}} = \mathbf{0} \quad \text{on} \quad \Gamma_{in}^0 \; \text{and} \; \Gamma_{wall}^0,
\end{equation}
\begin{equation}
\mathbf{n} \cdot \hat{\mathbf{T}} \cdot \mathbf{n} = 0 \quad \text{and} \quad \left(\mathbf{I} - \mathbf{n} \otimes \mathbf{n} \right) \cdot \hat{\mathbf{u}} = 0
\quad \text{on} \quad \Gamma_{out}^0,
\end{equation}
where the tensor $\mathbf{\hat{T}}$ is expressed by
\begin{align*}
\hat{\mathbf{T}} = -\hat{p} + 2 Nf^{-1} \hat{\mathbf{E}} \left( \hat{\mathbf{u}} \right); \quad \quad
\hat{\mathbf{E}} \left( \hat{\mathbf{u}} \right) = \frac{1}{2} \left[ \nabla \hat{\mathbf{u}} +  {\nabla \hat{\mathbf{u}}}^T \right].
\end{align*} 
We stress that while it is customary to impose additional conditions along the axis of symmetry $\Gamma_{sym}^0$, we did not apply any such conditions in this case because the model equations were written around the perturbed three-dimensional domain and then linearised before integrating out the $\theta$ dependence.

\subsection{Numerical method and validation}
\label{sec:linear_stability_solution}
\subsubsection{Linear algebra}
The linear stability of the steady state solutions as a parametric function of the system dimensionless groups is determined by solving a generalized, asymmetric matrix eigenvalue problem given by equation \eqref{eq:ls_algebraic_sys_eqns0}. The asymmetric nature of the problem, as can be seen by inspection of the Jacobian matrix, is due to the convective and interface deformation contributions to the matrix. In addition, it can also be seen that the mass matrix is singular, with several rows having identically zero entries, which can be attributed to the  absence of time derivative terms in the continuity equation and  though not obvious, due to the requirement that the perturbation vectors satisfy homogeneous essential boundary conditions \citep{carvalho_scriven_1999,Natarajan_1992}.
The implication of having singularity in the mass matrix is that the number of true eigenvalues is smaller than the dimension of the problem, with the number of missing eigenvalues equal to the number of algebraic equations having identically zero row in the matrix \citep{carvalho_scriven_1999}. We carry out a shift-and-invert transformation \citep{Christodoulou_scriven_1988} to map these missing, or so-called `infinite', eigenvalues to zero wherein the original problem is transformed to
\begin{align}
\LB \mathbf{J} - \upsilon \mathbf{B} \RB^{-1} \mathbf{B} \mathbf{y} = \tau   \mathbf{y},  \label{eq:ls_simple_eigenvalue_prob} 
\end{align}
where the shift $\upsilon$ is a complex constant, $\tau$ is the eigenvalue of the new problem and is related to the eigenvalue of the original problem $\beta$ by
\begin{equation}
\tau =  \frac{1}{\LB \beta - \upsilon \RB}. 
\end{equation}
We solve this eigenvalue problem by using an iterative Arnoldi method available in ARPACK \citep{Lehoucq_etal_1997}, which can be called within FreeFem++, using the standard \textit{Taylor-Hood} element for the flow field variables  and piecewise quadratic element for interface deformation magnitude as in the steady state simulations. The accuracy of the converged leading eigenvalue is confirmed by ensuring that the residual 
$\left | \mathbf{J} \mathbf{y} - \beta \mathbf{B} \mathbf{y} \right |$ 
is always less than $1 \times 10^{-10}$.
%
%
%
%
\subsubsection{Validation}	\label{sec:linear_stability_validation}
We test the validity of our theoretical and numerical procedure by examining the stability of a spherical bubble of fixed volume in a stagnant liquid with negligible gravitational and boundary effects. 
The bubble is stable under these conditions and its motion is governed by an analytical solution  \citep{Prosperetti_1980,Miller_Scriven_1968}. We compare our numerical results for the eigenvalues with this solution given in \cite{Prosperetti_1980} for  small amplitude normal mode perturbations. The characteristic scales used for the non-dimensionalisation of space, velocity, and pressure in the governing equations are $R$, $\sqrt{\gamma/(\rho R)}$, and $\gamma/R$, respectively, where $R$ is the bubble radius, so that the validation problem is parameterised by the Ohnesorge number,  $ Oh \mu/\sqrt{\rho R \gamma}$.


Based on the scaling above, the dimensionless form of the characteristic equation for the bubble oscillations reads \citep{Prosperetti_1980} 
\begin{align}
&\left[ \mathcal{H}^{\LB 1 \RB }_{m-\frac{1}{2}}\left( X^* \right) \right]\beta^2 + Oh\left[4m(m+2)^2 - 2(m+2)(2m+1)(\mathcal{H}^{(1)}_{m-\frac{1}{2}}\left( X^*\right) + 2)\right]\beta  \nonumber \\
&+ (m+1)(m-1)(m+2)(\mathcal{H}^{(1)}_{m-\frac{1}{2}}\left( X^*\right) + 2) = 0,  \label{eq:ls_validation_characteristic_eqn}
\end{align}
where $X^*$ is a rescaled growth rate, and $\mathcal{H}^{\LB 1 \RB }_{j}(X^*)$ is a Hankel function of the first kind: 
\begin{equation}
X^* = \left[ \frac{\beta}{Oh}\right]^{\frac{1}{2}}, ~~ {\rm and} ~~~
\mathcal{H}^{(1)}_j(X^*) = \frac{X^* H^{(1)}_{j+1}(X^*)}{H^{(1)}_j(X^*)}. 
\end{equation}
For a fixed value of $Oh$, we solve iteratively for $\beta$. The initial guess used is the solution to the following equation \citep{Prosperetti_1980}  
\begin{equation}
\beta^2  -  2Oh\left[(m+2)(2m+1) \right]\beta   + (m+1)(m-1)(m+2) = 0.
\end{equation}
Once the solution for the first eigenvalue is obtained, we use the associated $X^*$ for the previous $Oh$ as the initial guess for the next value of $Oh$. We implemented the solution steps in MATLAB and generated the analytical solution for $0 \le Oh \le 1$.

At steady state, in the absence of gravity and since the liquid surrounding the bubble is stagnant $(\mathbf{u} = 0)$, the governing equations reduce to
\begin{align}
\nabla p &= 0,  \label{eq:ls_validation_gov_eqn} \intertext{and the normal stress boundary condition to}		
-p + P_b  &=  \kappa \; \text{on} \; {\Gamma}_b. 
\end{align}
Equation \eqref{eq:ls_validation_gov_eqn} implies that pressure field in the liquid phase surrounding the bubble is a constant, $P_a$, so that the bubble pressure becomes
\begin{equation}
P_b = \kappa + P_a  \;~~~ \text{on} \; {\Gamma}_b.
\end{equation}
For the linear stability analysis, the value of $P_a$ was set to zero without loss of generality.
%

\begin{figure}
\centering
    \begin{subfigure}{0.48\linewidth}
        \includegraphics[width = \linewidth]{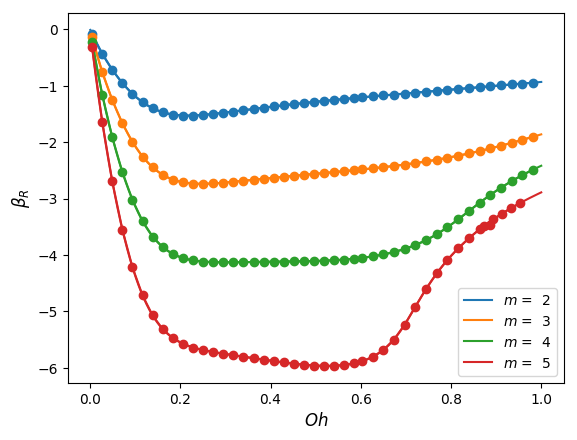}
	    \caption{}
	    \label{fig:ls_stability_validation_results_real} 
	\end{subfigure}
    \begin{subfigure}{0.48\linewidth}
        \includegraphics[width = \linewidth]{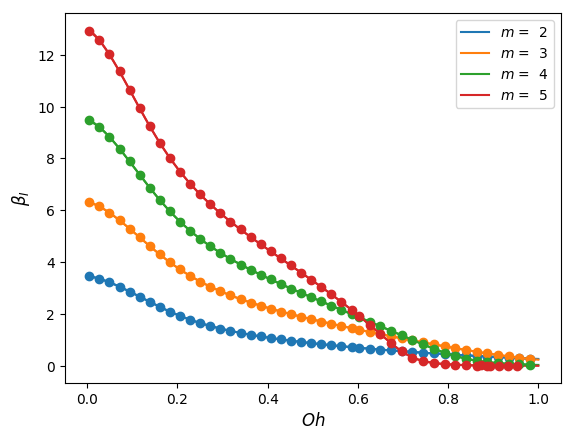}
	    \caption{}
	    \label{fig:ls_stability_validation_results_imag} 
    \end{subfigure}
    \caption{Validation of the theoretical and numerical procedure for an oscillating  bubble in the absence of gravitational and boundary effects. Comparison between the amplification rates, (a), and oscillation frequencies, (b), from the analytical solution given by equation \eqref{eq:ls_validation_characteristic_eqn} (coloured continuous solid line) and our numerically-generated growth rates (coloured markers), for modes m = 2, 3, 4 and 5.} 
\end{figure}

We solve the modified forms of the perturbation equations \eqref{eq:ls_weak_form_normal_mode_perturbation_radial}-\eqref{eq:ls_weak_form_normal_mode_perturbation_kinematic_bc} using the base state solutions computed as set out above. Figures \ref{fig:ls_stability_validation_results_real} and \ref{fig:ls_stability_validation_results_imag} respectively show excellent agreement between  the real and imaginary parts of the eigenvalues computed and the analytical solution of \eqref{eq:ls_validation_characteristic_eqn} as a function of the Ohnesorge number for four different azimuthal wavenumbers.

%
%
%
\section{Linear stability results}
 \label{sec:linear_stability_results}
In this section, we provide a discussion of the linear stability results starting with the dependence of the growth rate $\beta_R$ obtained from the leading eigenvalues associated primarily with the first two modes, $m=1$ and $m=2$, 
as a parametric function of $Nf$, $Eo$, and $U_m$. The asymmetric bubble shape near instability onset, associated with the most dangerous linear mode, is also discussed, and a stability map is  plotted which clearly demarcates the stability boundary as a function of the system parameters. 
 %
 %
\subsection{Dominant modes of instability}
 \label{sec:growth_rate}
\begin{figure} 
    \centering
    \begin{subfigure}[h]{0.32\linewidth}
        \includegraphics[width=\linewidth]{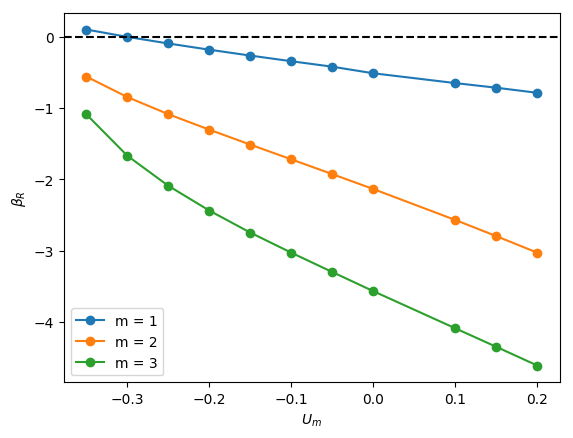}
        \caption{}
         \label{fig:ls_stabilityNf40Eo20L3}
    \end{subfigure}
    \begin{subfigure}[h]{0.32\linewidth}
        \includegraphics[width=\linewidth]{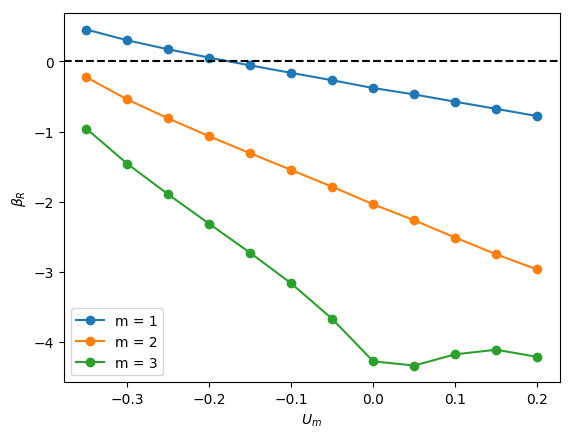}
        \caption{}
         \label{fig:ls_stabilityNf40Eo180L3}
    \end{subfigure}
    \begin{subfigure}[h]{0.32\linewidth}
        \includegraphics[width=\linewidth]{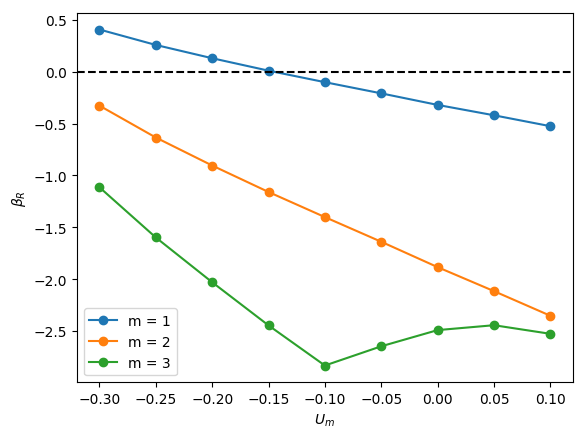}
        \caption{}
         \label{fig:ls_stabilityNf40Eo300L3}
    \end{subfigure}
    \begin{subfigure}[h]{0.32\linewidth}
        \includegraphics[width=\linewidth]{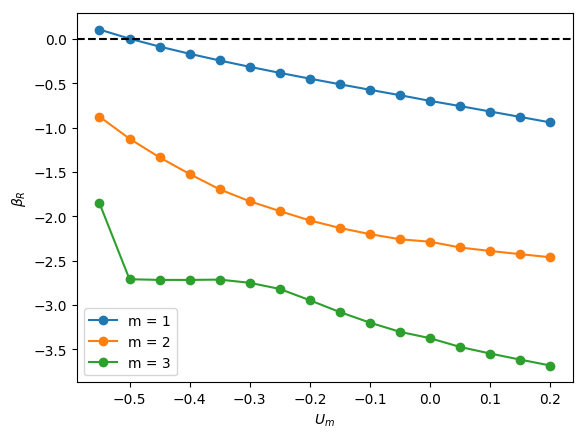}
        \caption{ }
        \label{fig:ls_stabilityNf60Eo20L3}
    \end{subfigure}
    \begin{subfigure}[h]{0.32\linewidth}
        \includegraphics[width=\linewidth]{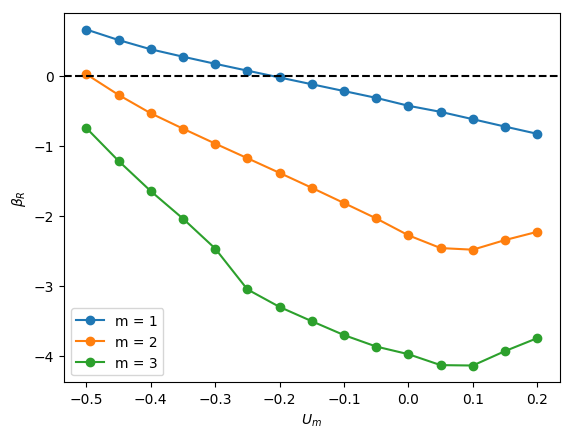}
        \caption{}
         \label{fig:ls_stabilityNf60Eo180L3}
    \end{subfigure}
    \begin{subfigure}[h]{0.32\linewidth}
        \includegraphics[width=\linewidth]{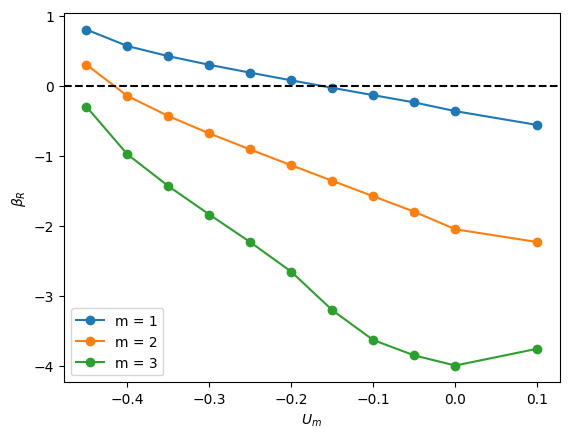}
        \caption{}
         \label{fig:ls_stabilityNf60Eo300L3}
    \end{subfigure}
    \centering
    \begin{subfigure}[h]{0.32\linewidth}
        \includegraphics[width=\linewidth]{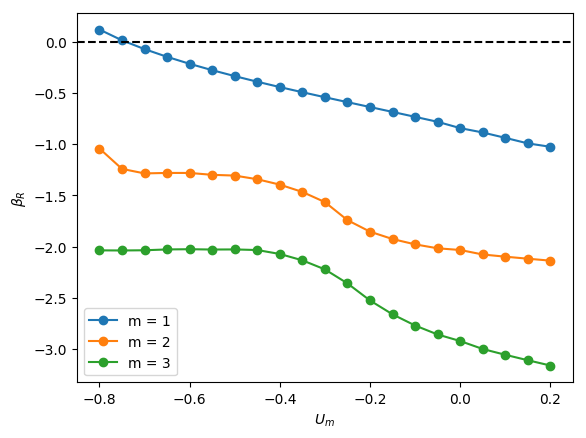}
        \caption{ }
        \label{fig:ls_stabilityNf80Eo20L3}
    \end{subfigure}
    \begin{subfigure}[h]{0.32\linewidth}
        \includegraphics[width=\linewidth]{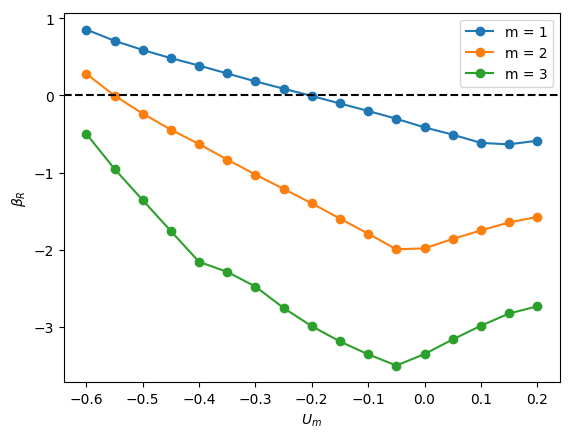}
        \caption{}
         \label{fig:ls_stabilityNf80Eo180L3}
    \end{subfigure}
    \begin{subfigure}[h]{0.32\linewidth}
        \includegraphics[width=\linewidth]{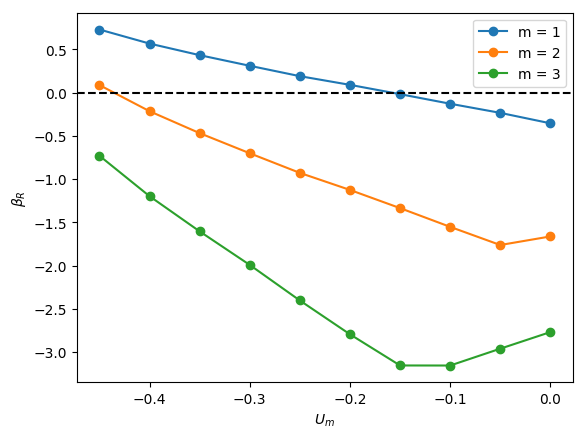}
        \caption{}
         \label{fig:ls_stabilityNf80Eo300L3}
    \end{subfigure}
    \centering
    \begin{subfigure}[h]{0.32\linewidth}
        \includegraphics[width=\linewidth]{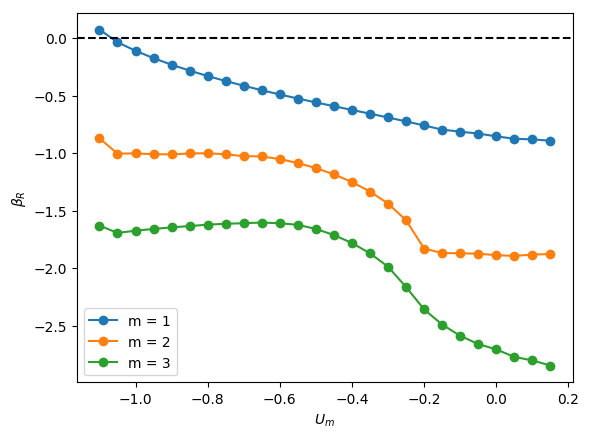}
        \caption{ }
        \label{fig:ls_stabilityNf100Eo20L3}
    \end{subfigure}
    \begin{subfigure}[h]{0.32\linewidth}
        \includegraphics[width=\linewidth]{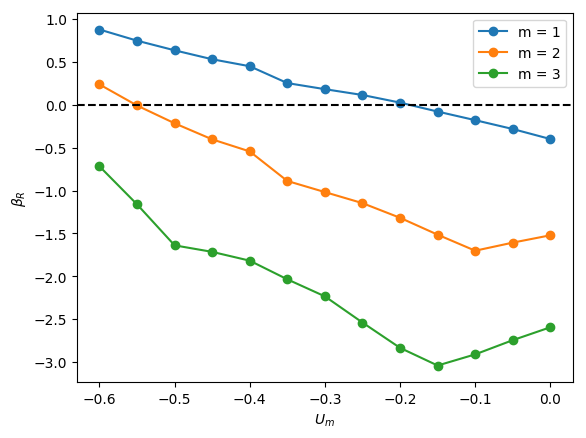}
        \caption{}
         \label{fig:ls_stabilityNf100Eo180L3}
    \end{subfigure}
    \begin{subfigure}[h]{0.32\linewidth}
        \includegraphics[width=\linewidth]{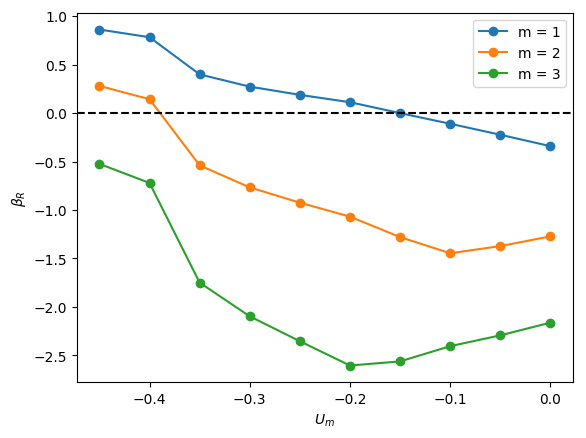}
        \caption{}
         \label{fig:ls_stabilityNf100Eo300L3}
    \end{subfigure}
     \caption{Growth rate, $\beta_R$, as a function of $U_m$ for $Nf = 40$, 60, 80, and 100, shown in (a)-(c), (d)-(f),  (g)-(i), and, (j)-(l), respectively, with $Eo = 20$ in (a), (d), (g), and (j), $Eo=180$ in (b), (e), (h), and (k), and $Eo=300$ in (c), (f), (i), and (l). The results are shown for the modes $m = 1, 2$, and $3$.} 
     \label{fig:ls_stabilityNf406080100}
\end{figure}

The linear stability of axisymmetric steady states associated with $40 \leq Nf \leq 100$ and $20 \leq Eo \leq 300$ was investigated for downward liquid flow characterised by $U_m <0$; these base states had been computed and reported in the companion paper to the present work by \cite{Abubakar_Matar_2021_s}. In Figure \ref{fig:ls_stabilityNf406080100}, we show the dependence of the growth rate $\beta_R$ on $U_m$ for modes $m=1$, $m=2$, and $m=3$ for $Nf=40,60,80,100$, and $Eo=20,180,300$. 
For all the cases shown in this figure, it is evident that $m=1$ is the most unstable mode, which corresponds to a deflection of the bubble away from the axis of symmetry and occurs over a well-defined range of negative $U_m$ values, $\Delta U_m$, for which $\beta_R >0$ indicating the presence of a linear instability. For sufficiently large $Eo$ and $Nf$, the $m=2$ is also unstable, though it remains sub-dominant to the $m=1$ mode, as illustrated in Figure \ref{fig:ls_stabilityNf406080100}(l), for instance, for the $Nf=100$ and $Eo=300$ case. Modes associated with $m=0$ and $m \geq 3$ have $\beta_R \leq 0$ for all values of $Nf$, $Eo$, and $U_m$ studied, and play no role in the transition to linear instability. Furthermore, for all the cases examined, the eigenvalues associated with the $m=1$ and $m=2$ modes are real.  

\begin{figure} 
    \centering
    \begin{subfigure}[h]{0.48\linewidth}
        \includegraphics[width=\linewidth]{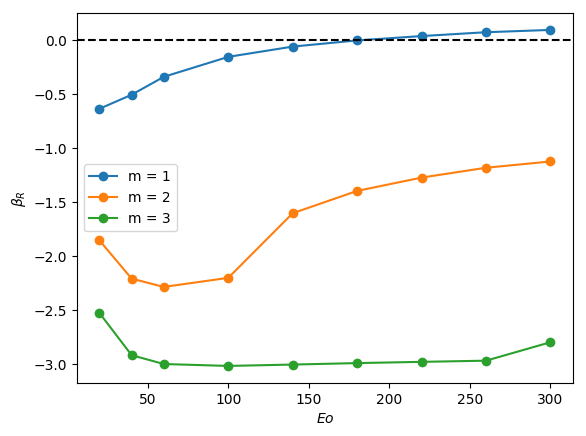}
        \caption{}
         \label{fig:ls_stabilityNf80UL20L3}
    \end{subfigure}
   \begin{subfigure}[h]{0.48\linewidth}
        \includegraphics[width=\linewidth]{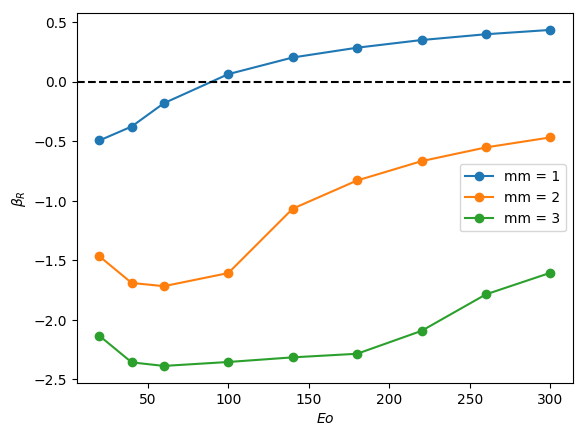}
        \caption{}
         \label{fig:ls_stabilityNf80UL35L3}
    \end{subfigure}
   \begin{subfigure}[h]{0.48\linewidth}
        \includegraphics[width=\linewidth]{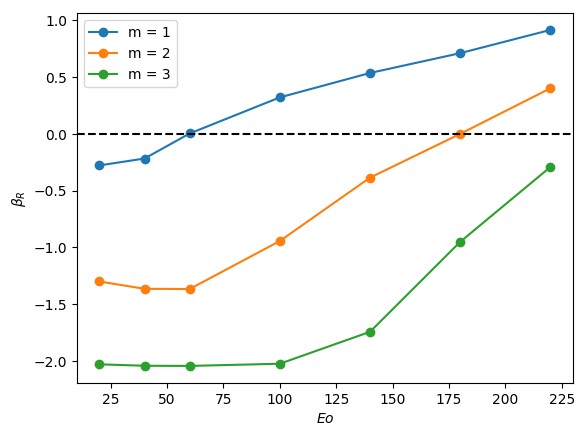}
        \caption{}
         \label{fig:ls_stabilityNf80UL55L3}
    \end{subfigure}
     \caption{Growth rate, $\beta_R$, as a function of $Eo$, with $Nf =80$ and  $U_m = -0.20$, (a), $U_m = -0.35$, (b) and $U_m = -0.55$, (c). The results are shown for $m = 1, 2$, and $3$.} 
     \label{fig:ls_stabilityNf80UL}
\end{figure}

From Figure \ref{fig:ls_stabilityNf406080100}, it is seen that for $Nf=40,60,80,100$, increasing $Eo$ is accompanied by a decrease in the magnitude of $U_m$ required for instability and a widening of $\Delta U_m$ though this trend appears to saturate at large $Eo$. Moreover, in Figure \ref{fig:ls_stabilityNf80UL}, which depicts the variation of $\beta_R$ with $Eo$ and with $Nf=80$ held constant, the critical $Eo$ for which the $m=1$ mode is destabilised is reduced threefold as $U_m$ is varied from -0.2 to -0.55. Thus, the results presented in Figures \ref{fig:ls_stabilityNf406080100} and \ref{fig:ls_stabilityNf80UL} demonstrate that 
increasing the velocity of the downward-flowing liquid and/or decreasing the relative significance of surface tension forces is destabilising. 

Inspection of Figure \ref{fig:ls_stabilityNf406080100} also reveals that decreasing $Nf$ for $Eo=180$ and $Eo=300$, that is, for weak surface tension, appears to have little effect on the critical $U_m$,  $\Delta U_m$, and the magnitude of $\beta_R$ for the most dangerous mode, $m=1$. In contrast, for $Eo=20$, decreasing $Nf$ leads to a substantial decrease in the critical $U_m$ value and is therefore strongly destabilising indicating that viscous effects gain in significance as the relative importance of surface tension increases with decreasing $Eo$ for sufficiently low $Eo$. Furthermore, from Figure \ref{fig:ls_stabilityNf406080100} (d,a), (e,b) and (f,c), it is also seen that decreasing $Nf$ from $Nf=60$ to $Nf=40$ also leads to a large reduction in the critical $U_m$ even at high $Eo$ values for sufficiently small $Nf$. 


We offer an explanation of the trends highlighted above by focusing on the bubble nose and appealing to the steady state results of \cite{Abubakar_Matar_2021_s} who had noted that the  frontal radius of curvature of the nose, $R_F$, in stagnant and  downward-flowing liquids increases with $Eo$ then saturates for $Eo \gtrsim 100$ and is weakly-dependent on $Nf$ for $Nf \gtrsim 60$. For $Nf < 60 $, $R_F$ is reduced with decreasing $Nf$. For $Eo < 100$, $R_F$ exhibits a turning point in $Eo$ for all $Nf$ values studied with a well-defined cross-over $Eo$ value below which the magnitude of the $R_F$ minima increase with decreasing $Nf$. These results demonstrate that bubble noses become flatter with decreasing and increasing $Nf$ for sufficiently small and large $Eo$, respectively. \cite{Abubakar_Matar_2021_s} have also shown that increasingly negative $U_m$ has a similar effect leading to flatter bubble noses regardless of the value of $Nf$ and $Eo$; this is attributed to the increase in the normal stress exerted on the bubble nose relative to that in a stagnant liquid due to the commensurate increase in the opposing inertial force in the downward liquid flow.
The results of \cite{Abubakar_Matar_2021_s} for the nose curvature and its dependence on $Nf$, $Eo$, and $Um$ thus appear to mirror the linear stability trends presented in Figures \ref{fig:ls_stabilityNf406080100}-\ref{fig:ls_stabilityNf80UL}. In particular, the significant destabilisation of the bubble with $Eo$ increasing from 20 to 180 and its subsequent saturation, the destabilising effect of $Nf$ with decreasing $Nf$ at $Eo=20$ (see figure 4(a,j)), the weak dependence on the stability characteristics for $Nf=40,60,80,100$  for $Eo=180,300$ (see Figure 4(b,k) and (c,l)) can be  correlated to the parametric dependence of the nose curvature $R_F$ on $Nf$, $Eo$, and $U_m$.
%
%


\subsection{Asymmetric bubble shapes}
 \label{sec:asymmetric_bubble_shapes}
We now study the influence of the parameters $Nf$, $Eo$, and $U_m$ on the shapes of the eigenfunctions  focusing on those associated with the interfacial deformation in order to highlight the base state bubble regions targeted by the instability. 
For every point on the three-dimensional axisymmetric base state interface with position vector $ \left( r^0, \theta^0, z^0 \right)$ in cylindrical coordinates,
from \eqref{eq:ls_interface_perturbation} 
and \eqref{eq:ls_nomal_mode_perturbation_forms_h}, and recalling that $\tilde{\mathbf{x}} = \tilde{h} \mathbf{n}^0$,  the corresponding deformed interface points in Cartesian coordinates can be constructed:
\begin{subequations}
\begin{align}
x &= r^0 \cos \left( \theta^0 \right) + \epsilon \left[ h_R n_r \cos \left( m \theta^0\right) - h_I n_r \sin \left( m \theta^0\right)\right] \cos \left( \theta^0\right) \quad \forall \; \theta^0 \; \in \; \left[0,2\pi \right]   \\
 y &= r^0 \sin \left( \theta^0 \right) + \epsilon \left[ h_R n_r \cos \left( m \theta^0 \right) - h_I n_r \sin \left( m \theta^0 \right)\right] \sin \left( \theta^0 \right) \quad \forall \; \theta^0 \; \in \; \left[0,2\pi \right]  \\
 z &= z^0 + \epsilon \left[ h_R n_z \cos \left( m \theta^0 \right) - h_I n_z \sin \left( m \theta^0 \right)\right] \quad \forall \; \theta^0 \; \in \; \left[0,2\pi \right] 
\end{align}
\end{subequations}
where $h_R$ and  $h_I$ denote the real and imaginary parts of the interface deformation in the normal direction; $n_r$ and $n_z$  remain the radial and axial components of the unit normal to the base state interface, respectively. 
In our discussion below of the three-dimensional bubble shape immediately following the transition to instability, we assign a value to the parameter $\epsilon$, which signifies the formally infinitesimal size of the perturbation, to enhance the visualisation of the results. 
 
\begin{figure} 
    \centering
    \begin{subfigure}[h]{0.32\linewidth}
        \includegraphics[width=\linewidth]{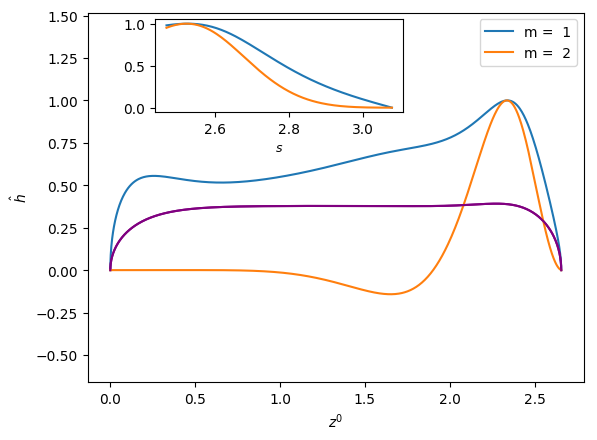}
        \caption{}
         \label{fig:ls_axissym2DBubbleNf40Eo20UL20L3}
    \end{subfigure}
    \begin{subfigure}[h]{0.32\linewidth}
        \includegraphics[width=\linewidth]{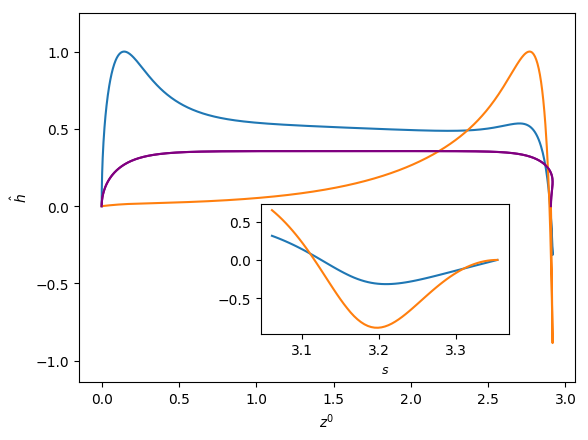}
        \caption{}
         \label{fig:ls_axissym2DBubbleNf40Eo180UL20L3}
    \end{subfigure}
    \begin{subfigure}[h]{0.32\linewidth}
        \includegraphics[width=\linewidth]{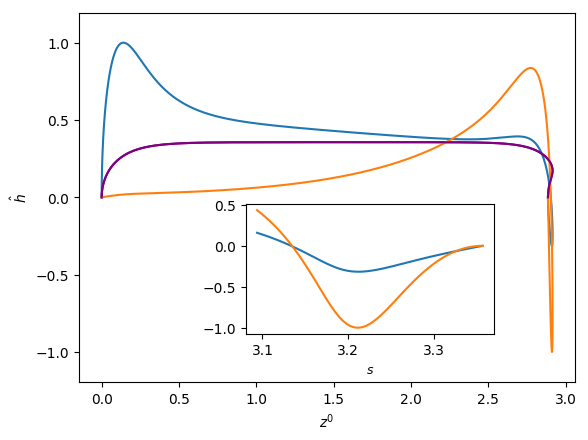}
        \caption{}
         \label{fig:ls_axissym2DBubbleNf40Eo260UL20L3}
    \end{subfigure}
    \begin{subfigure}[h]{0.32\linewidth}
        \includegraphics[width=\linewidth]{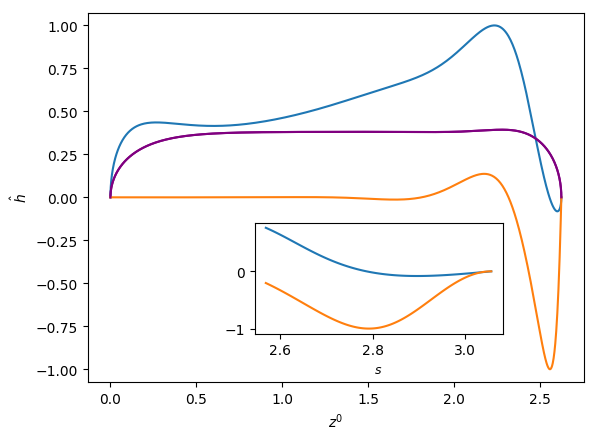}
        \caption{ }
        \label{fig:ls_axissym2DBubbleNf60Eo20UL20L3}
    \end{subfigure}
    \begin{subfigure}[h]{0.32\linewidth}
        \includegraphics[width=\linewidth]{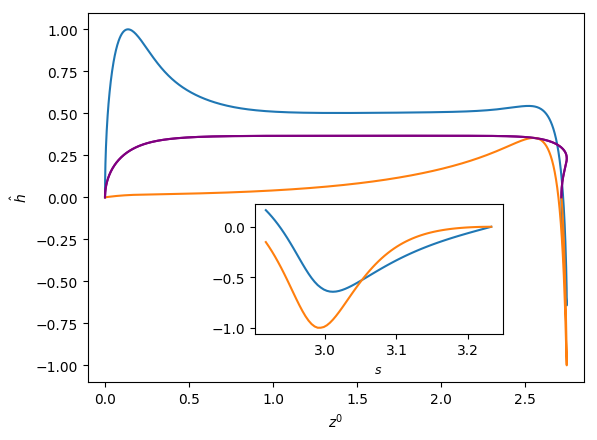}
        \caption{}
         \label{fig:ls_axissym2DBubbleNf60Eo180UL20L3}
    \end{subfigure}
    \begin{subfigure}[h]{0.32\linewidth}
        \includegraphics[width=\linewidth]{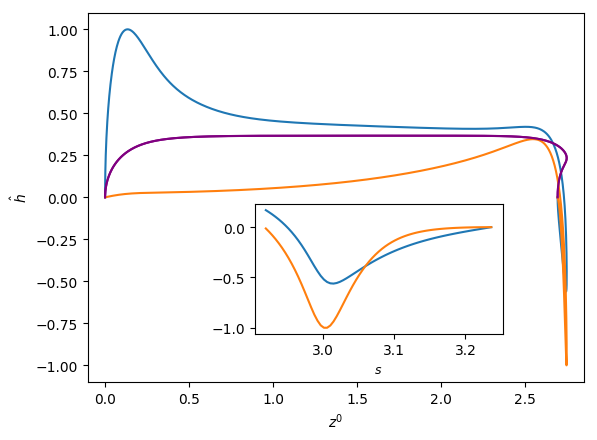}
        \caption{}
         \label{fig:ls_axissym2DBubbleNf60Eo260UL20L3}
    \end{subfigure}
    \centering
    \begin{subfigure}[h]{0.32\linewidth}
        \includegraphics[width=\linewidth]{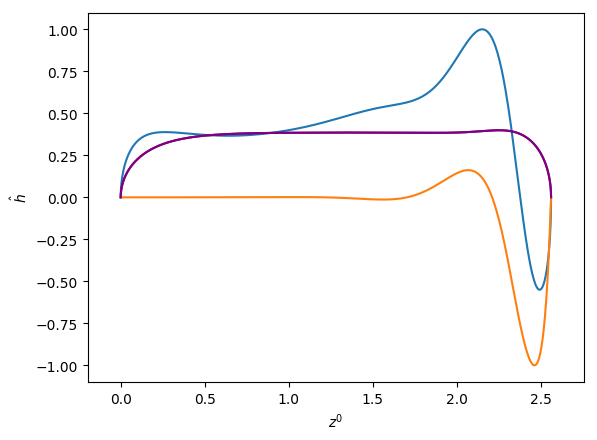}
        \caption{ }
        \label{fig:ls_axissym2DBubbleNf80Eo20UL20L3}
    \end{subfigure}
    \begin{subfigure}[h]{0.32\linewidth}
        \includegraphics[width=\linewidth]{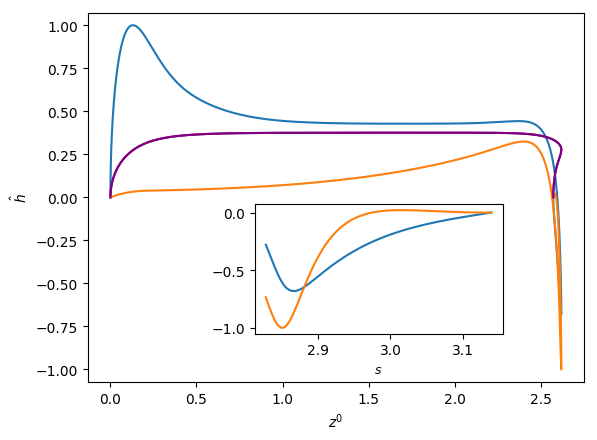}
        \caption{}
         \label{fig:ls_axissym2DBubbleNf80Eo180UL20L3}
    \end{subfigure}
    \begin{subfigure}[h]{0.32\linewidth}
        \includegraphics[width=\linewidth]{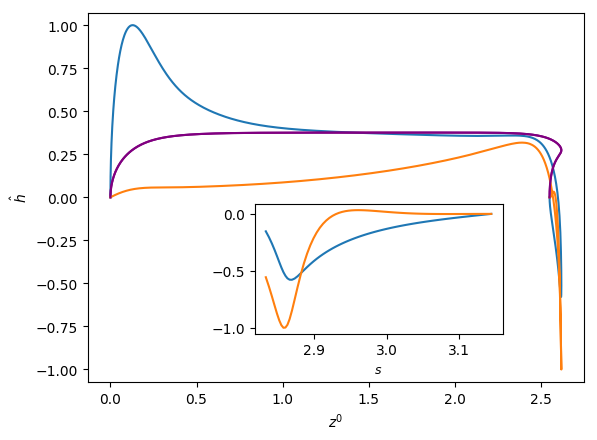}
        \caption{}
         \label{fig:ls_axissym2DBubbleNf80Eo260UL20L3}
    \end{subfigure}
    \centering
    \begin{subfigure}[h]{0.32\linewidth}
        \includegraphics[width=\linewidth]{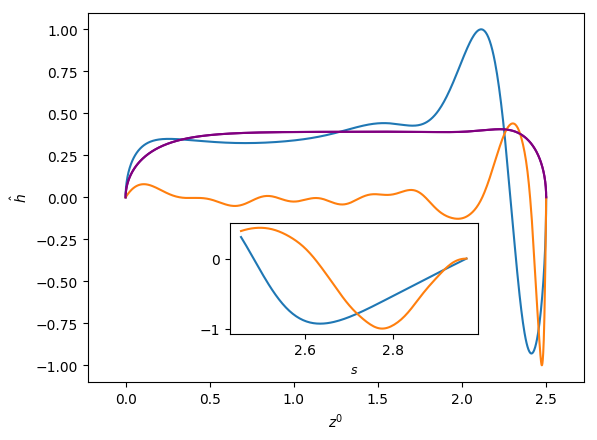}
        \caption{ }
        \label{fig:ls_axissym2DBubbleNf100Eo20UL20L3}
    \end{subfigure}
    \begin{subfigure}[h]{0.32\linewidth}
        \includegraphics[width=\linewidth]{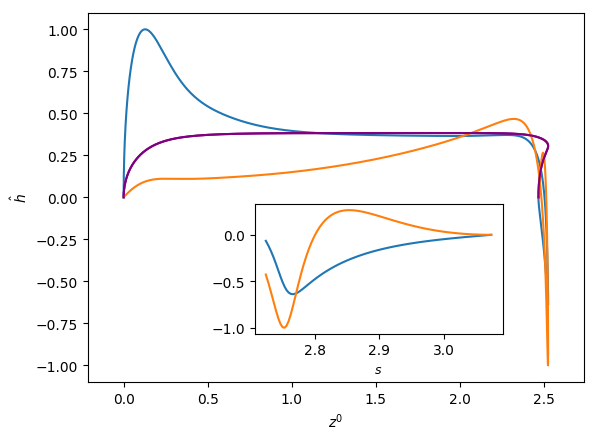}
        \caption{}
         \label{fig:ls_axissym2DBubbleNf100Eo180UL20L3}
    \end{subfigure}
    \begin{subfigure}[h]{0.32\linewidth}
        \includegraphics[width=\linewidth]{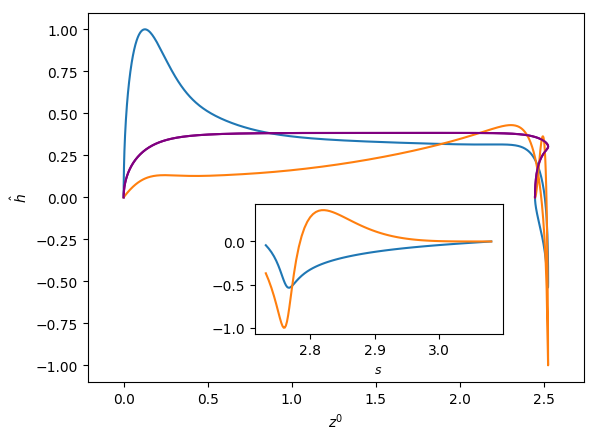}
        \caption{}
         \label{fig:ls_axissym2DBubbleNf100Eo260UL20L3}
    \end{subfigure}
     \caption{Interface deformation eigenfunctions $\hat{h}$ for eigenmodes $m=1$ and $m=2$ as a function of base state axial position of the interface, $z^0$, with $U_m = -0.20$ and $Nf = 40$, 60, 80, and 100, shown in (a)-(c), (d)-(f),  (g)-(i), and (j)-(l), respectively, with $Eo = 20$ in (a), (d), (g), and (j), $Eo=180$ in (b), (e), (h), and (k), and $Eo=300$ in (c), (f), (i), and (l). The axisymmetric base state bubble shape is also shown (coloured purple) as a reference in order to highlight the regions targeted by the instability. The insets depict enlarged views of $\hat{h}$ varying with the arc length $s$ for $m=1$ and $m=2$ in the bubble bottom region.} 
     \label{fig:ls_axissym2DBubbleNf406080100}
\end{figure}

\begin{figure} 
    \centering
    \begin{subfigure}[h]{0.48\linewidth}
        \includegraphics[width=\linewidth]{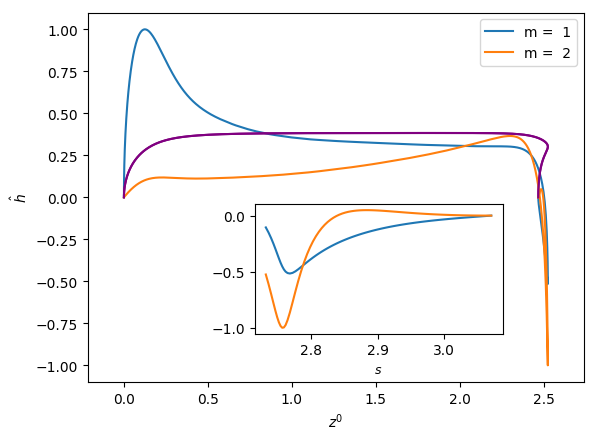}
        \caption{}
         \label{fig:ls_axissym2DBubbleNf100Eo220UL25L3}
    \end{subfigure}
    \begin{subfigure}[h]{0.48\linewidth}
        \includegraphics[width=\linewidth]{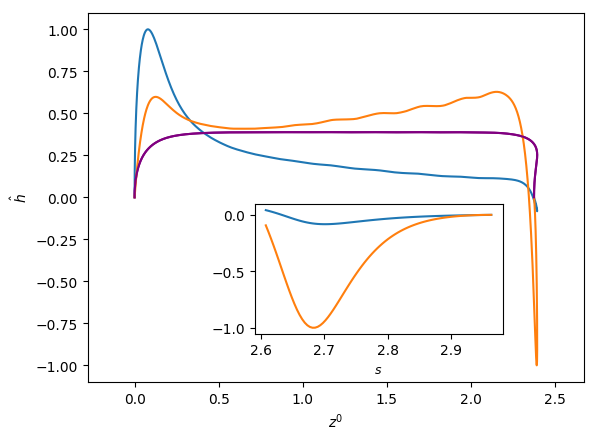}
        \caption{}
         \label{fig:ls_axissym2DBubbleNf100Eo220UL40L3}
    \end{subfigure}
    \begin{subfigure}[h]{0.48\linewidth}
        \includegraphics[width=\linewidth]{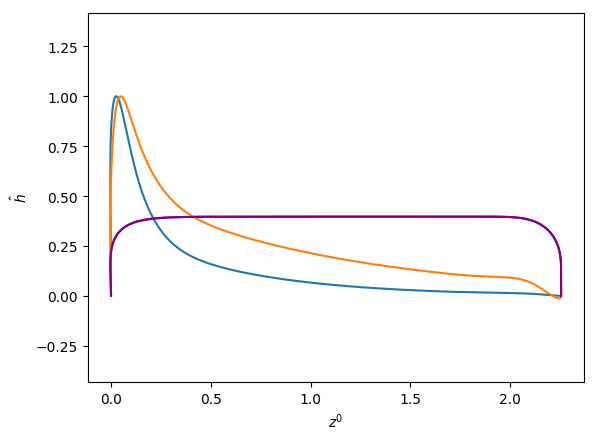}
        \caption{}
       \label{fig:ls_axissym2DBubbleNf100Eo220UL55L3}
    \end{subfigure}
     \caption{Interface deformation eigenfunctions $\hat{h}$ for eigenmodes $m=1$ and $m=2$ as a function of base state axial position of the interface, $z^0$, for $Nf = 100$, $Eo=220$, and with $U_m = -0.25$, (a), $U_m = -0.40$, (b), and  $U_m = -0.55$, (c). The axisymmetric base state bubble shape is also shown (coloured purple) in order to highlight the regions targeted by the instability. The insets depict enlarged views of $\hat{h}$ varying with the arc length $s$ for $m=1$ and $m=2$ in the bubble bottom region.}
      \label{fig:ls_axissym2DBubbleNf100UL254055}
\end{figure}

 Figures \ref{fig:ls_axissym2DBubbleNf406080100} and \ref{fig:ls_axissym2DBubbleNf100UL254055} show the influence of the parameters $Nf$, $Eo$ and $U_m$ on the eigenfunctions for the interface deformation $\hat{h}$. It is seen that the  the nose, film, and bottom regions of the bubble are targeted to varying degrees; the precise definitions of these regions are in \cite{Abubakar_Matar_2021_s}. For the dominant eigenmode $m = 1$, the peaks in $\hat{h}$ coincide with the nose and bottom regions for high and low $Eo$, respectively.  
Around the bottom region, the observed peaks in $\hat{h}$ are either due to the tail structure at high $Eo$ (see the middle and right column in Figure 6), or the undulation in the film region close to the bubble bottom at low $Eo$ (see the left column in Figure 6) though in the latter case we note that the $m=1$ mode is linearly stable for the parameters used to generate these results ($U_m=-0.2$, $Eo=20$, and $Nf=40,60,80,100$). For eigenmode $m = 2$, the peak in $\hat{h}$ coincides with the bottom region except at higher magnitude of downward liquid flow velocity, $U_m$, where similar peaks are seen in the nose region, as shown in  Figure \ref{fig:ls_axissym2DBubbleNf100Eo220UL55L3}. In Figures  \ref{fig:ls_axissym2DBubbleNf406080100} and \ref{fig:ls_axissym2DBubbleNf100UL254055}, we also show enlarged views of the variation of $\hat{h}$ with the arc length $s$ for the bottom region to demonstrate that these boundary layer-like regions in $\hat{h}$ have been resolved adequately.


\begin{figure} 
    \centering
    \begin{subfigure}[h]{0.35\linewidth}
        \includegraphics[width=\linewidth]{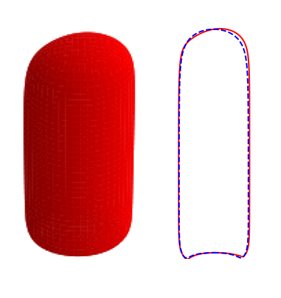}
        \caption{}
         \label{fig:ls_fullBubbleNf100Eo220UL25L3M1Angle120}
    \end{subfigure}
    \begin{subfigure}[h]{0.35\linewidth}
        \includegraphics[width=\linewidth]{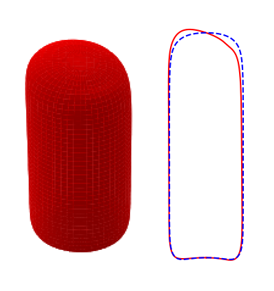}
        \caption{}
         \label{fig:ls_fullBubbleNf100Eo220UL40L3M1Angle0}
    \end{subfigure}
    \
    \begin{subfigure}[h]{0.35\linewidth}
        \includegraphics[width=\linewidth]{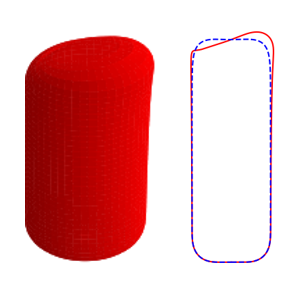}
        \caption{}
         \label{fig:ls_fullBubbleNf100Eo220UL55L3M1Angle60}
    \end{subfigure}
    \begin{subfigure}[h]{0.35\linewidth}
        \includegraphics[width=\linewidth]{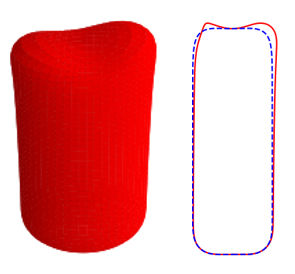}
        \caption{}
         \label{fig:ls_fullBubbleNf100Eo220UL55L3M2Angle150}
    \end{subfigure}
     \caption{Three-dimensional bubble shapes and their two-dimensional projections (solid lines) in the $(r,z)$ plane obtained by adding the interface deformation for $m=1$ to the base state (dashed lines) with $\epsilon=0.05$ for $U_m = -0.25, -0.40, -0.55$, shown in (a)-(c), respectively, with $Nf = 100$ and $Eo = 220$. In (d), we show the analogous shape for $m=2$ with $U_m=-0.55$, $Nf=100$, and $Eo=220$.} 
     \label{fig:ls_fullBubbleNf100Eo220ULM1}
\end{figure}

\begin{figure} 
    \centering
    \begin{subfigure}[h]{0.35\linewidth}
        \includegraphics[width=\linewidth]{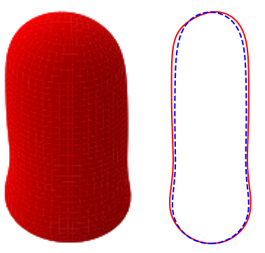}
        \caption{}
         \label{fig:ls_fullBubbleNf40Eo20UL35L3M1Angle0}
    \end{subfigure}
    \begin{subfigure}[h]{0.35\linewidth}
        \includegraphics[width=\linewidth]{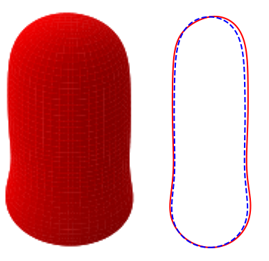}
        \caption{}
         \label{fig:ls_fullBubbleNf60Eo20UL55L3M1Angle0}
    \end{subfigure}
    \begin{subfigure}[h]{0.35\linewidth}
        \includegraphics[width=\linewidth]{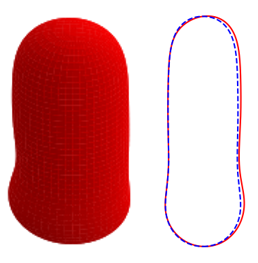}
        \caption{}
         \label{fig:ls_fullBubbleNf80Eo20UL80L3M1Angle0}
    \end{subfigure}
    \begin{subfigure}[h]{0.35\linewidth}
        \includegraphics[width=\linewidth]{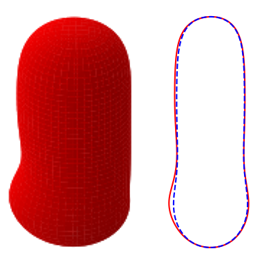}
        \caption{}
         \label{fig:ls_fullBubbleNf100Eo20UL110L3M1Angle30}
    \end{subfigure}
     \caption{\textcolor{black}{Three-dimensional bubble shapes and their two-dimensional projections (solid lines) in the $(r,z)$ plane obtained by adding the interface deformation for $m=1$ to the base state (dashed lines) with $\epsilon=0.05$ for $U_m = -0.35, -0.55, -0.80, -1.10$, and $Nf = 40, 60, 80, 100$, shown in (a)-(d), respectively, with $Eo = 20$.}} 
     \label{fig:ls_fullBubbleNfEo20ULM1}
\end{figure}

In Figures \ref{fig:ls_fullBubbleNf100Eo220ULM1} and \ref{fig:ls_fullBubbleNfEo20ULM1},
we show the effects of $U_m$, $Eo$, and $Nf$ on the three-dimensional bubble shapes obtained by adding the interface deformation  associated with mode $m = 1$ to the base state taking the value of $\epsilon = 0.05$ for clarity of presentation; these correspond to the shapes one might expect to observe experimentally at the onset of instability. Also shown in Figures \ref{fig:ls_fullBubbleNf100Eo220ULM1} and \ref{fig:ls_fullBubbleNfEo20ULM1} are two-dimensional projections of the shapes in the $(r,z)$ plane which highlight the deviations from the base state within the framework of linear theory. The results shown in Figure \ref{fig:ls_fullBubbleNf100Eo220ULM1} are for the same parameter values used to generate Figure \ref{fig:ls_axissym2DBubbleNf100UL254055}, which correspond to the large $Eo$, weak surface tension limit; panel (d) of this figure also depicts the analogous results for the deformed bubble shape when $m = 2$. These results illustrate the asymmetry of the nose for Taylor bubble motion in downward flowing liquids for negligible surface tension and are reminiscent of the asymmetric shape shown in Figure \ref{fig:asym_bubble_fabre}. In Figure \ref{fig:ls_fullBubbleNfEo20ULM1}, on the other hand, the results are associated with $Eo=20$ at which surface tension effects are significant. Here, it is clearly seen that in addition to asymmetries in the nose region, the instability also targets the undulation in the bottom region. Figure \ref{fig:ls_fullBubbleNf100Eo20UL110L3M1Angle30} provides a clear demonstration that the asymmetry is most pronounced in this region for the fastest downward flowing liquid case; in contrast, the bubble nose remains essentially axisymmetric in this case. 


%

%
%
%
 %
 %
 \subsection{Stability maps}
 \label{sec:instability_map}
\begin{figure} 
    \centering
    \begin{subfigure}[h]{0.48\linewidth}
        \includegraphics[width=\linewidth]{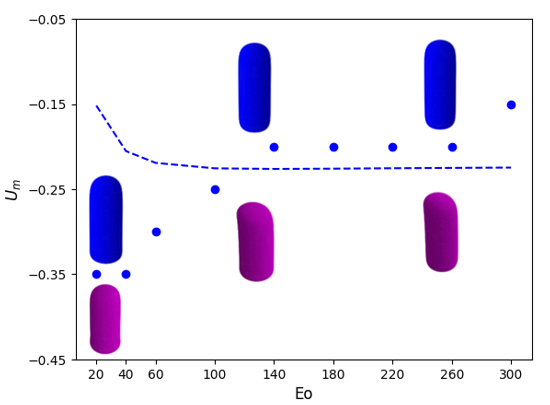}
        \caption{}
         \label{fig:ls_regimeStabilityNf40}
    \end{subfigure}
    \begin{subfigure}[h]{0.48\linewidth}
        \includegraphics[width=\linewidth]{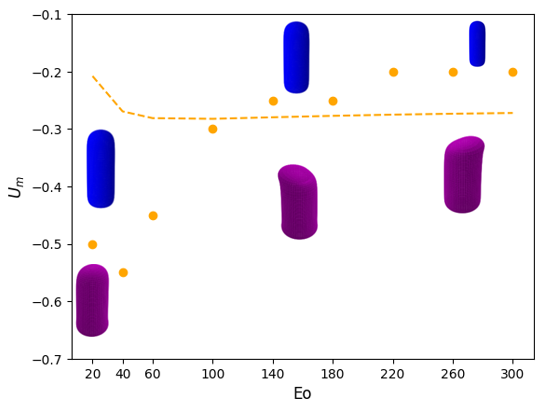}
        \caption{}
         \label{fig:ls_regimeStabilityNf60}
    \end{subfigure}
    \begin{subfigure}[h]{0.48\linewidth}
        \includegraphics[width=\linewidth]{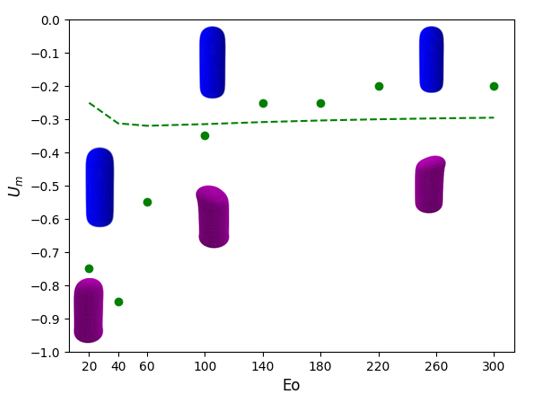}
        \caption{}
         \label{fig:ls_regimeStabilityNf80}
    \end{subfigure}
    \begin{subfigure}[h]{0.48\linewidth}
        \includegraphics[width=\linewidth]{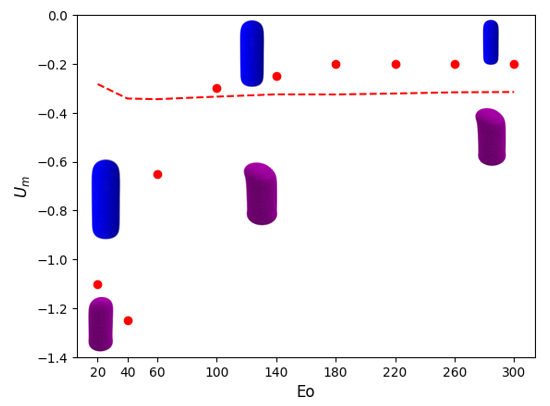}
        \caption{}
         \label{fig:ls_regimeStabilityNf100}
    \end{subfigure}
    \begin{subfigure}[h]{0.48\linewidth}
        \includegraphics[width=\linewidth]{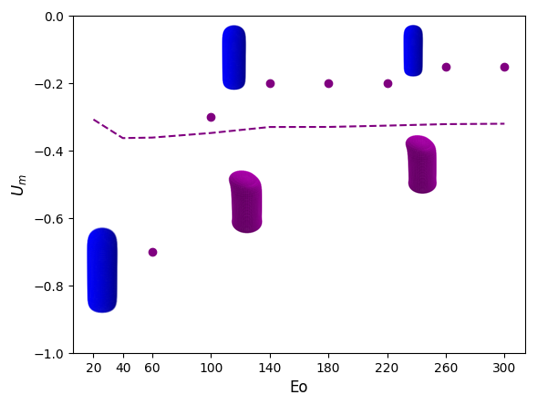}
        \caption{}
         \label{fig:ls_regimeStabilityNf120}
    \end{subfigure}
     \caption{Stability maps depicting the boundaries demarcating (shown by the full circles) the regions of linear instability in $(U_m,Eo)$ space characterised by a transition from axisymmetric to asymmetric bubble shapes in downward liquid flow with $U_m <0$ for (a) $Nf = 40$, (b) $Nf = 60$, (c) $Nf = 80$, (d) $Nf = 100$ (e) $Nf = 120$. The dashed lines represent the curves of  critical $U_m$ for which the axisymmetric base state corresponds to an arrested bubble in downward liquid flow with $U_b = 0$.} 
     \label{fig:ls_regimeStabilityNf}
\end{figure}

 We show in Figure \ref{fig:ls_regimeStabilityNf} stability maps in $(U_m,Eo)$ with $Nf$ varying parametrically that depict  the boundaries  demarcating regions of linear instability for Taylor bubbles moving in downward flowing liquids characterised by $U_m < 0$. In each case, examples of the three-dimensional, asymmetric bubble shape at the onset of instability is also shown. 
The general trend observed is that for $Eo>60$, the magnitude of the critical $U_m$ decreases with increasing $Eo$ for a fixed $Nf$, saturating for large $Eo$ and $Nf$, beyond $Eo \gtrsim 100$ and $Nf \gtrsim 60$. For $Eo<60$, there is a turning point in the stability map that highlights the fact that an increase in the magnitude of the downward flow velocity is needed to overcome the strong surface tension forces at sufficiently low $Eo$; furthermore, the critical $U_m$ magnitude increases with $Nf$ reflecting the destabilising effect of viscous stresses over this range of $Eo$. The results presented in Figure \ref{fig:ls_regimeStabilityNf} are consistent with the trends discussed in the previous sections, and, in particular,  those associated with the normalised frontal radius of curvature of the nose, $R_F$, presented in \cite{Abubakar_Matar_2021_s}; this provides a further indication that the dependence of $R_F$ on $U_m$, $Eo$, and $Nf$ controls the linear stability characteristics of the bubble motion. 

We also show in Figure \ref{fig:ls_regimeStabilityNf} the curve for the critical $U_m$ for which the axisymmetric base state has $U_b = 0$. It is noticeable that for
$Eo \gtrsim 100$ for all $Nf$ studied, this curve is in the linearly unstable region indicating that bubbles whose motion has been arrested due to a downward flowing liquid over this range of $Eo$ and $Nf$ cannot have an axisymmetric shape. In contrast, in the complementary range of $Eo$, the critical $U_m$ curve for such bubbles is outside the unstable region implying that they can sustain an axisymmetric shape despite the downward liquid flow. 
Examples of these cases have been observed experimentally; see, for instance,  the axisymmetric Taylor bubble shown to have been held stationary in downward liquid flow by \cite{Nigmatulin_2001}, for $Nf = 6087.38$ and $Eo = 33.06$. We note that the $Eo$ value is within the range of the linearly stable region shown in Figure 10(e). Though the experimental $Nf$ value is outside of the range we studied, the fact that the linear stability boundaries appear to saturate at large $Nf$ suggests that our results can still provide a reasonable  indication of the behaviour observed experimentally.  
%
%
%
%
\section{Energy budget analysis} 
\label{sec:linear_stability_energy_budget}
In this section, we analyse how energy is  transferred from the base flow to the perturbations by studying the growth of the perturbation kinetic energy \citep{Hu_Joseph_1989,Hooper_Boyd_1983}. By investigating the contribution of the mechanisms of different physical origin that account for energy production, one can identify the dominant ones that drive instability \citep{Boomkamp_Miesen_1996}. This analysis has been used to study destabilising mechanisms in parallel two-phase flows \citep{Sahu_etal_2009,Selvam_etal_2007,Sahu_etal_2007,Onarai_etal_2011} and their classification \citep{Boomkamp_Miesen_1996}.

\subsection{Energy balance formulation}
\label{sec:energy_balance_model}
To derive the equation for the growth of the disturbance kinetic energy, one multiplies the continuous forms of the momentum perturbation equations for the velocity components with their corresponding complex conjugates, integrates over the domain, adds the resulting equations, and simplifies as appropriate. Our perturbation equations for the velocity components, however, were derived from the weak form of the continuous momentum equations written around the perturbed domain; the development of the weak form of the energy equation must therefore follow the same strategy. Thus, we obtain the energy equation from the derived perturbation equations for the velocity components by setting the test functions for the latter to equal the complex conjugate of the velocity perturbations, 
$\bar{{\Phi}}_r = \hat{{u}}_r^*$, 
$\bar{{\Phi}}_\theta = \hat{{u}}_\theta^*$, 
$\bar{{\Phi}}_z = \hat{{u}}_z^*$,
followed by necessary simplifications: 
\begin{align}
 \beta_R &  
\int_{\Omega^0} 
\left \{
\left| \hat{u}_r \right|^2 + \left| \hat{u}_\theta \right|^2 + \left| \hat{u}_z \right|^2
\right \}  d \Omega^0 
+
\int_{\Omega^0} 
\left \{
\left| \hat{u}_r \right|^2 \frac{\partial u_r^0}{\partial r^0} + \left| \hat{u}_z \right|^2 \frac{\partial u_z^0}{\partial z} + \frac{\left| \hat{u}_\theta \right|^2 u_r^0}{r} 
\right. \nonumber
\\ &\left.
+ 
\left( \frac{\partial u_r^0}{\partial z} + \frac{\partial u_z^0}{\partial r^0} \right) \mathbb{R}\left \{ \hat{u}_r \hat{u}_z^*\right \}
\right \}  d {\Omega^0}  
+ 
\int_{\Omega}
Nf^{-1}
\left \{
2\left| \frac{\partial \hat{u}_r}{\partial r}\right|^2 + 2\left| \frac{\partial \hat{u}_z}{\partial z}\right|^2
\right. \nonumber
\\ &\left.
+ 
\frac{1}{r^2} 
\left[ 
\left( 2 + m^2 \right)\left| \hat{u}_r \right|^2 + \left( 1 + 2m^2 \right)\left| \hat{u}_\theta \right|^2  + m^2 \left|\hat{u}_z \right|^2
\right] 
+
\left| \frac{\partial \hat{u}_\theta}{\partial r}\right|^2
+
\left| \frac{\partial \hat{u}_\theta}{\partial z}\right|^2
\right. \nonumber
\\ &\left.
+
\left| \frac{\partial \hat{u}_r}{\partial z} + \frac{\partial \hat{u}_z}{\partial r}\right|^2  
- 
\frac{1}{r} \frac{\partial \left| \hat{u}_\theta \right|^2}{\partial r} 
+ \frac{6m}{{r}^2} \mathbb{I}m \left \{ \hat{u}_r \hat{u}_\theta^*\right \}
+ \frac{2m}{r} \mathbb{I}m \left( \hat{u}_z^* \frac{\partial \hat{u}_\theta}{\partial z} + \hat{u}_r^* \frac{\partial \hat{u}_\theta}{\partial r}  \right)
\right \}
d {\Omega^0} \nonumber
\\
&-\int_{\Gamma_b^0}
Eo^{-1}
\left \{
-\frac{d \hat{h}}{d s} \left[ \mathbf{n} \cdot \frac{d \hat{\mathbf{u}}^*}{ds} - \kappa_a \left( \mathbf{t} \cdot \hat{\mathbf{u}}^* \right)\right]
+
\hat{h} \left[\kappa_a^2 + \kappa_b^2 - \frac{m^2}{r^2} \right] \mathbf{n} \cdot \hat{\mathbf{u}}^* 
\right \} 
d\Gamma_b^0 \nonumber
\\
&-
\int_{\Gamma_b^0}
\left\{ 
\hat{h} n_z 
\LB
 \mathbf{n} \cdot \hat{\mathbf{u}}^*
 \RB
\right\}d\Gamma_b^0 \nonumber
\\
&+ 
\int_{\Gamma^0_b}
\hat{h}
\left\{
{
\LB
\mathbf{u}^0 \cdot \mathbf{t}
\RB
\left[ 
\frac{d \mathbf{u}^0}{d s} \cdot \hat{\mathbf{u}}^*
\right] 
}
+
\LS -p^0 + 2 Nf^{-1} \LB \mathbf{t} \cdot \frac{d \mathbf{u}^0}{d s} \RB  \RS \LB \mathbf{t} \cdot \frac{d \hat{\mathbf{u}}^*}{d s} \RB  \right. \nonumber 
\\ &\left.
+
\LS -p^0 + 2 Nf^{-1} \frac{u_r^0}{r}  \RS \LB \frac{\hat{u}_r^*}{r} - \mathfrak{i} m  \frac{\hat{u}_\theta^*}{r}\RB 
\right\}
d\Gamma^0_b  \nonumber
\\
&+
\int_{\Gamma_b^0}
\left\{ 
\left[ 
{Eo}^{-1} \kappa  + z - P_b^0
\right] 
\LS 
\LB \mathbf{t} \cdot \hat{\mathbf{u}}^* \RB \frac{d \hat{h}}{ds} + \hat{h} \LB  \mathfrak{i} m  \frac{\hat{u}_\theta^*}{r} \RB 
+
\hat{h} \kappa \LB \mathbf{n} \cdot \hat{\mathbf{u}}^* \RB
\RS
\right\}d\Gamma_b^0  = 0 \label{eq:ls_energy_balance_eqn}
\end{align}
where the symbol $|.|$ represents the magnitude of a complex function; $\mathbb{R}$ and $\mathbb{I}$ denote the real and imaginary part of a complex function, respectively. Equation \eqref{eq:ls_energy_balance_eqn} is the energy budget formulation that governs the evolution of the disturbance kinetic equation. 


\indent Following \cite{Boomkamp_Miesen_1996}, we express the energy balance equation as 
\begin{equation}
\dot{E} = REY + DIS + INT, \label{eq:Edot}
\end{equation}
where $\dot{E}$ corresponds to the time range of change of the perturbation kinetic energy given by the following relation
\begin{equation}
\dot{E} = \beta_R 
\int_{\Omega^0} 
\left \{
\left| \hat{u}_r \right|^2 + \left| \hat{u}_\theta \right|^2 + \left| \hat{u}_z \right|^2
\right \}  d \Omega^0\equiv \beta_R ~KIN, 
\end{equation}
wherein $KIN$ represents the total kinetic energy associated with the perturbation velocity field, which equals $\dot{E}$ when multiplied by the growth rate $\beta_R$ (which is positive for an unstable flow). We also introduce the following definitions for the terms $REY$ and $DIS$ that appear on the right-hand-side of equation (\ref{eq:Edot}) \citep{Boomkamp_Miesen_1996}: 
\begin{equation}
REY = 
-\int_{\Omega^0} 
\left \{
\left| \hat{u}_r \right|^2 \frac{\partial u_r^0}{\partial r^0} + \left| \hat{u}_z \right|^2 \frac{\partial u_z^0}{\partial z} + \frac{\left| \hat{u}_\theta \right|^2 u_r^0}{r} 
+ \left( \frac{\partial u_r^0}{\partial z} + \frac{\partial u_z^0}{\partial r^0} \right) \mathbb{R}\left \{ \hat{u}_r \hat{u}_z^*\right \}
\right \}  d {\Omega^0}, \\
\end{equation}
\begin{align}
DIS =
&-\int_{\Omega}
Nf^{-1}
\left \{
2\left| \frac{\partial \hat{u}_r}{\partial r}\right|^2 + 2\left| \frac{\partial \hat{u}_z}{\partial z}\right|^2
+ 
\frac{1}{r^2} 
\left[ 
\left( 2 + m^2 \right)\left| \hat{u}_r \right|^2 + \left( 1 + 2m^2 \right)\left| \hat{u}_\theta \right|^2  
\right. \right. 
\nonumber \\ 
&
\left. \left.
+ m^2 \left|\hat{u}_z \right|^2
\right] 
+
\left| \frac{\partial \hat{u}_\theta}{\partial r}\right|^2
+
\left| \frac{\partial \hat{u}_\theta}{\partial z}\right|^2
+
\left| \frac{\partial \hat{u}_r}{\partial z} + \frac{\partial \hat{u}_z}{\partial r}\right|^2  
- 
\frac{1}{r} \frac{\partial \left| \hat{u}_\theta \right|^2}{\partial r} 
+ \frac{6m}{{r}^2} \mathbb{I}m \left \{ \hat{u}_r \hat{u}_\theta^*\right \}
\right. 
\nonumber \\ 
&
\left.
+ \frac{2m}{r} \mathbb{I}m \left( \hat{u}_z^* \frac{\partial \hat{u}_\theta}{\partial z} + \hat{u}_r^* \frac{\partial \hat{u}_\theta}{\partial r}  \right)
\right \}
d {\Omega}. 
\end{align}
Here, $REY$ denotes the rate of energy transfer by the Reynolds stress from the base flow to the disturbed flow, and $DIS$ represents the rate of viscous dissipation of energy of the disturbed flow. We also provide a breakdown for $INT$, the rate of work done by the velocity and stress disturbances in deforming the interface \citep{Boomkamp_Miesen_1996}
\begin{subequations}
\begin{align}
&INT = NOR + TAN, \\
&NOR = TEN + HYD + BUB,
\end{align}
\end{subequations}
which we have decomposed into its normal, $NOR$, and tangential, $TAN$, components with $NOR$ 
further subdivided into $TEN$, $HYD$, and $BUB$, representing work done at the interface against surface tension, gravity, and bubble pressure, respectively. Expressions for $TAN$, $NOR$, and $TEN$, $HYD$ and $BUB$, are respectively provided by
%
%
\begin{align}
TAN 
=
&
\mathbb{R}
\left \{
-\int_{\Gamma^0_b}
\hat{h}
\left\{
{
\LB
\mathbf{u}^0 \cdot \mathbf{t}
\RB
\left[ 
\frac{d \mathbf{u}^0}{d s} \cdot \hat{\mathbf{u}}^*
\right] 
}
+
\LS -p^0 + 2 Nf^{-1} \LB \mathbf{t} \cdot \frac{d \mathbf{u}^0}{d s} \RB  \RS \LB \mathbf{t} \cdot \frac{d \hat{\mathbf{u}}^*}{d s} \RB  
\right. 
\right.  \nonumber 
\\ 
&
\left. 
\left.
+
\LS -p^0 + 2 Nf^{-1} \frac{u_r^0}{r}  \RS \LB \frac{\hat{u}_r^*}{r} - \mathfrak{i} m  \frac{\hat{u}_\theta^*}{r}\RB 
\right\}
d\Gamma^0_b 
+
\int_{\Gamma_b^0}
Eo^{-1}
\left \{
\frac{d \hat{h}}{d s} \left[  \kappa_a \left( \mathbf{t} \cdot \hat{\mathbf{u}}^* \right)\right]
\right \} 
d\Gamma_b^0 
\right. 
  \nonumber 
\\ 
&
\left.
-\int_{\Gamma_b^0}
\left\{ 
\left[ 
{Eo}^{-1} \kappa  + z - P_b^0
\right] 
\LS 
\LB \mathbf{t} \cdot \hat{\mathbf{u}}^* \RB \frac{d \hat{h}}{ds} + \hat{h} \LB  \mathfrak{i} m  \frac{\hat{u}_\theta^*}{r} \RB 
\RS
\right\}d\Gamma_b^0
\right \}, \label{eq:ls_energy_balance_interface_tan}
\end{align}
\begin{align}
NOR = 
&
\mathbb{R}
\left \{
\int_{\Gamma_b^0}
Eo^{-1}
\left \{
-\frac{d \hat{h}}{d s} \left[ \mathbf{n} \cdot \frac{d \hat{\mathbf{u}}^*}{ds} \right]
+
\hat{h} \left[\kappa_a^2 + \kappa_b^2 - \frac{m^2}{r^2} \right] \mathbf{n} \cdot \hat{\mathbf{u}}^* 
\right \} 
d\Gamma_b^0 
\right. \nonumber \\
&
\left.
+
\int_{\Gamma_b^0}
\left\{ 
\hat{h} n_z 
\LB
 \mathbf{n} \cdot \hat{\mathbf{u}}^*
 \RB
\right\}d\Gamma_b^0 
-\int_{\Gamma_b^0}
\left\{ 
\left[ 
{Eo}^{-1} \kappa  + z - P_b^0
\right] 
\LS 
\hat{h} \kappa \LB \mathbf{n} \cdot \hat{\mathbf{u}}^* \RB
\RS
\right\}d\Gamma_b^0
\right \}, 
\end{align}
\begin{equation}
TEN = 
\mathbb{R}
\left \{
\int_{\Gamma_b^0}
Eo^{-1}
\left \{
-\frac{d \hat{h}}{d s} \left[ \mathbf{n} \cdot \frac{d \hat{\mathbf{u}}^*}{ds} \right]
+
\hat{h} \left[\kappa_a^2 + \kappa_b^2 - \kappa^2 - \frac{m^2}{r^2} \right] \mathbf{n} \cdot \hat{\mathbf{u}}^* 
\right \} 
d\Gamma_b^0 
\right \},
\end{equation}
\begin{equation}
HYD = 
\mathbb{R}
\left \{
\int_{\Gamma_b^0}
\left\{ 
\hat{h} \LB n_z - z \kappa \RB 
\LB
 \mathbf{n} \cdot \hat{\mathbf{u}}^*
 \RB
\right\}d\Gamma_b^0 
\right \},
\end{equation}
\begin{equation}
BUB = 
\mathbb{R}
\left \{
\int_{\Gamma_b^0}
\left\{ 
P_b^0 
\LS 
\hat{h} \kappa \LB \mathbf{n} \cdot \hat{\mathbf{u}}^* \RB
\RS
\right\}d\Gamma_b^0
\right \}.
\end{equation}

We normalised  the energy terms using $KIN$, so  the energy balance equation becomes
\begin{equation}
\beta_R = REY^* + DIS^* + TEN^* + HYD^* + BUB^* + TAN^*, \label{eq:ls_energy_balance_eqn_scaled}
\end{equation}  
where the asterisk designates the normalisation by $KIN$. 
In Table \ref{tab:ls_energy_balance}, we demonstrate  for eigenmode $m=1$, $Nf = 80$, and $Eo = 140$, and various $U_m$ values, that the difference between the growth rate $\beta_R$ on the left-hand-side of \eqref{eq:ls_energy_balance_eqn_scaled} computed from the linear stability analysis and the sum of the energy terms on the right-hand-side of \eqref{eq:ls_energy_balance_eqn_scaled}, $SUM$, denoted as $DIF$ in Table \ref{tab:ls_energy_balance} is negligibly small. These results inspire confidence in our procedure for computing the terms in \eqref{eq:ls_energy_balance_eqn_scaled}.
%
\begin{table}
\centering
	\caption{Balance of energy distribution among the various energy terms in equation \eqref{eq:ls_energy_balance_eqn_scaled} for $Nf = 80$, $Eo = 140$, and mode $m = 1$.}
  \label{tab:ls_energy_balance}
\begin{tabular}{cccccccccc} 
\hline
$U_m$&$\beta_R$&$REY^*$&$DIS^*$&$TEN^*$&$HYD^*$&$BUB^*$&$TAN^*$&$SUM$&$DIF$ \\  
$-0.40$&$0.2824$&$-0.0059$&$-2.6795$&$-0.1070$&$0.3066$&$5.9845$&$-3.21631$&$0.2824$&$2.78 \times 10^{-8}$ \\ 
$-0.30$&$0.1128$&$0.1224$&$-2.6494$&$0.0528$&$0.5397$&$3.1162$&$-1.0690$&$0.1128$&$1.85 \times 10^{-7}$ \\ 
$-0.20$&$-0.6354$&$0.2656$&$-2.6537$&$0.2919$&$0.8280$&$0.9208$&$0.2840$&$-0.6354$&$4.2.0 \times 10^{-8}$ \\ 
$-0.10$&$-0.2442$&$0.3260$&$-2.6008$&$0.5245$&$1.2070$&$-0.5126$&$0.8117$&$-0.2442$&$2.23 \times 10^{-6}$ \\ 
$0.00$&$-0.4474$&$0.2269$&$-2.1745$&$0.5412$&$1.1340$&$-0.9108$&$0.7358$&$-0.4473$&$1.40 \times 10^{-5}$ \\ 
$0.10$&$-0.6404$&$0.0413$&$-1.6299$&$0.4210$&$0.6429$&$-0.5639$&$0.4481$&$-0.6404$&$-7.67 \times 10^{-6}$ \\
\end{tabular}
\end{table} 
%
 \begin{figure} 
    \centering
    \begin{subfigure}[h]{0.48\linewidth}
        \includegraphics[width=\linewidth]{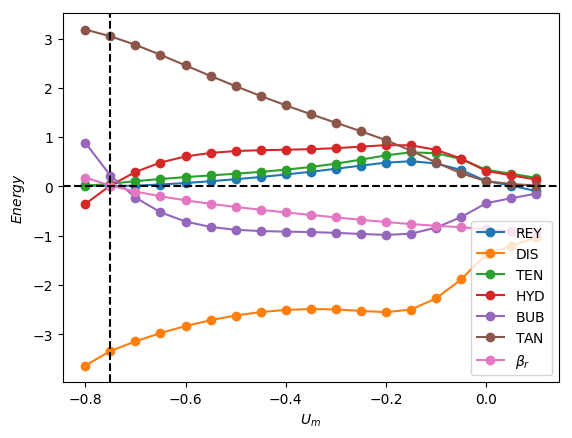}
        \caption{ }
        \label{fig:ls_energyNf80Eo20L3M1}
    \end{subfigure}
    \begin{subfigure}[h]{0.48\linewidth}
        \includegraphics[width=\linewidth]{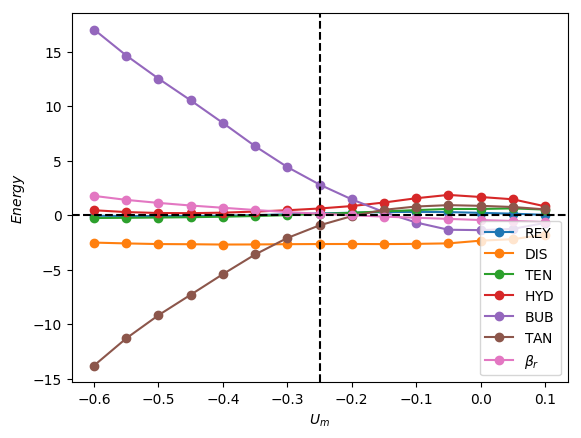}
        \caption{}
         \label{fig:ls_energyNf80Eo180L3M1}
    \end{subfigure}
    \begin{subfigure}[h]{0.48\linewidth}
        \includegraphics[width=\linewidth]{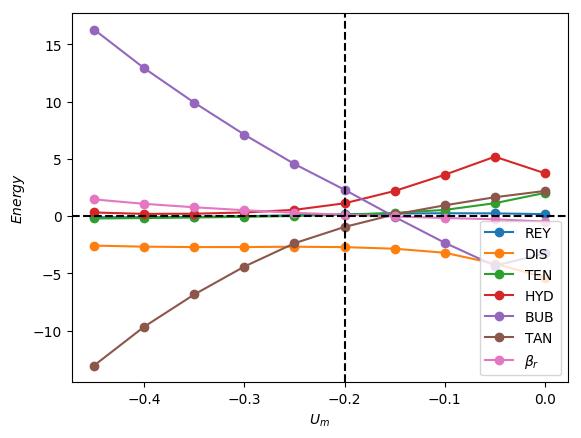}
        \caption{}
         \label{fig:ls_energyNf80Eo300L3M1}
    \end{subfigure}
     \caption{Breakdown of the energy budget for the eigenmode $m=1$ as a function of $U_m$ for $Eo=20$, 180, and 300, shown in (a)-(c), respectively, with $Nf = 80$. In each panel, the vertical dashed line marks the $U_m$ value for which $\beta_R=0$.} 
     \label{fig:ls_energyNf80EoL3M1}
\end{figure}
%
%
%
%
\subsection{Energy analysis results}
\label{sec:energy_analysis_of_ls_results}
From the linear stability analysis results for downward liquid flow presented in section \ref{sec:linear_stability_results}, the $m=1$ mode was identified as being the most unstable one. Here, we examine the contribution of each term in equation \eqref{eq:ls_energy_balance_eqn_scaled} in order to elucidate their roles in driving instability  and to identify the most dominant destabilising mechanism. 
In Figure \ref{fig:ls_energyNf80EoL3M1}, the energy analysis results for $Nf = 80$ with $Eo = 20, 180,$ and $300$ are shown. These parameter values are chosen to correspond to those used to generate a representative subset of the results presented in section \ref{sec:linear_stability_results}. In the discussion below, the asterisk decoration which appears in equation \eqref{eq:ls_energy_balance_eqn_scaled} is suppressed for the sake of brevity.  

It is seen clearly in Figure \ref{fig:ls_energyNf80Eo20L3M1} that for $Eo=20$, $TAN$ is overwhelmingly the dominant mechanism, followed by $BUB$, over the range of $U_m$ for which $\beta_R > 0$. Over the remainder of the $U_m$ range studied in which $\beta_R <0$, $TAN$ decreases monotonically but remains destabilising while $BUB$ becomes stabilising. $HYD$, $REY$, and $TEN$ are destabilising over this range while, as expected, $DIS$ is stabilising. Upon increasing $Eo$ to $180$ and $300$, as shown in Figure \ref{fig:ls_energyNf80Eo180L3M1} and \ref{fig:ls_energyNf80Eo300L3M1}, respectively, the dominant mechanism over the unstable range of $U_m$ switches to $BUB$ followed in relative dominance by $HYD$, which is marginally destabilising, while $TAN$ is strongly stabilising. These results suggest that the dominant mechanism depends on the relative significance of surface tension forces. In the case of negligible surface tension, characterised by high $Eo$ values, the destabilising mechanism is related to the bubble pressure which indicates that the origin of the instability is in the gas phase. 
At low $Eo$, the instability originates in the liquid phase with the energy provided by the tangential stress component.

It is instructive to split $TAN$, given by equation \eqref{eq:ls_energy_balance_interface_tan},  into its constituent components based on the base state groups that supply energy to the perturbations:
\begin{subequations} \label{eq:ls_energy_balance_interface_tan_comp}
\begin{align}
&TAN_{ut} = \mathbb{R}
\left \{
-\int_{\Gamma^0_b}
\hat{h}
{
\LB
\mathbf{u}^0 \cdot \mathbf{t}
\RB
\left[ 
\frac{d \mathbf{u}^0}{d s} \cdot \hat{\mathbf{u}}^*
\right] 
}
\right \}, \\
&TAN_{strs} 
=
\mathbb{R}
\left \{
-\int_{\Gamma^0_b}
\hat{h}
\left\{
\LS -p^0 + 2 Nf^{-1} \LB \mathbf{t} \cdot \frac{d \mathbf{u}^0}{d s} \RB  \RS \LB \mathbf{t} \cdot \frac{d \hat{\mathbf{u}}^*}{d s} \RB 
\right.
\right.
\nonumber
\\
&
\left.
\left.
\qquad {} \qquad {} \qquad {} 
+ \LS -p^0 + 2 Nf^{-1} \frac{u_r^0}{r}  \RS \LB \frac{\hat{u}_r^*}{r} - \mathfrak{i} m  \frac{\hat{u}_\theta^*}{r}\RB
\right \}
\right \}, \\
&TAN_{ts} = 
\mathbb{R}
\left \{
\int_{\Gamma_b^0}
Eo^{-1}
\left \{
\frac{d \hat{h}}{d s} \left[  \kappa_a \left( \mathbf{t} \cdot \hat{\mathbf{u}}^* \right)\right]
-
\kappa 
\LS 
\LB \mathbf{t} \cdot \hat{\mathbf{u}}^* \RB \frac{d \hat{h}}{ds} + \hat{h} \LB  \mathfrak{i} m  \frac{\hat{u}_\theta^*}{r} \RB 
\RS
\right \} 
d\Gamma_b^0
\right \}, \\
&TAN_g = 
\mathbb{R}
\left \{
\int_{\Gamma_b^0}
-z
\LS 
\LB \mathbf{t} \cdot \hat{\mathbf{u}}^* \RB \frac{d \hat{h}}{ds} + \hat{h} \LB  \mathfrak{i} m  \frac{\hat{u}_\theta^*}{r} \RB 
\RS
d\Gamma_b^0
\right \}, \\
&TAN_{pb} = 
\mathbb{R}
\left \{
\int_{\Gamma_b^0}
P_b
\LS 
\LB \mathbf{t} \cdot \hat{\mathbf{u}}^* \RB \frac{d \hat{h}}{ds} + \hat{h} \LB  \mathfrak{i} m  \frac{\hat{u}_\theta^*}{r} \RB 
\RS
d\Gamma_b^0
\right \}.
\end{align}
\end{subequations}
The terms $TAN_{ut}$, $TAN_{strs}$, $TAN_{ts}$, $TAN_{g}$ and $TAN_{pb}$ denote the contributions to $TAN$ due to streaming tangential velocity, tangential stress, surface tension, gravity and bubble pressure on the interface as captured by the base state terms  $\mathbf{u}  \cdot \mathbf{t}$, $\LS -p^0 + 2 Nf^{-1} \LB \mathbf{t} \cdot \frac{d \mathbf{u}^0}{d s} \RB  \RS + \LS -p^0 + 2 Nf^{-1} \frac{u_r^0}{r}  \RS $, 
$Eo$, $z$ and $P_b$ in the expressions, respectively. The base state contribution to $TAN_{strs}$ is related to the tangential stress because it can be obtained by taking the double dot product of the stress tensor and the tangential projection operator, $\left( \mathbf{I} - \mathbf{n} \otimes  \mathbf{n} \right)$. 

\begin{figure} 
    \centering
    \begin{subfigure}[h]{0.48\linewidth}
        \includegraphics[width=\linewidth]{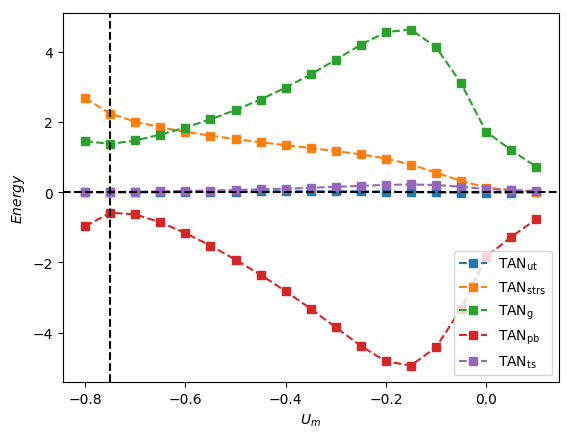}
        \caption{ }
        \label{fig:ls_tanEnergyNf80Eo20L3M1}
    \end{subfigure}
    \begin{subfigure}[h]{0.48\linewidth}
        \includegraphics[width=\linewidth]{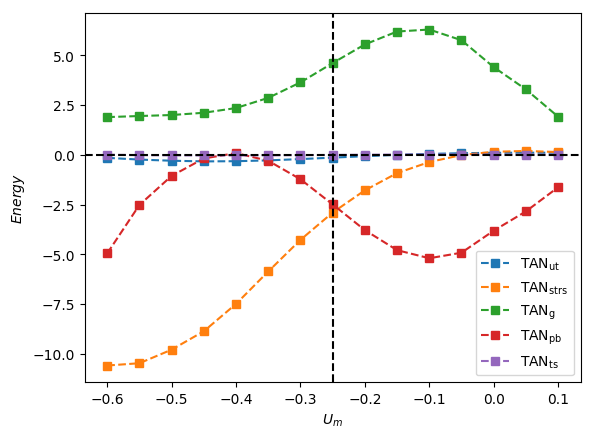}
        \caption{}
         \label{fig:ls_tanEnergyNf80Eo180L3M1}
    \end{subfigure}
    \begin{subfigure}[h]{0.48\linewidth}
        \includegraphics[width=\linewidth]{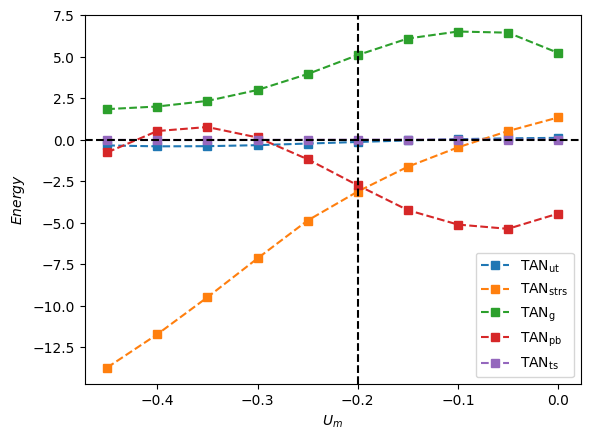}
        \caption{}
         \label{fig:ls_tanEnergyNf80Eo300L3M1}
    \end{subfigure}
     \caption{Breakdown of $TAN$ for the eigenmode $m=1$ into its constituent components given by equations \eqref{eq:ls_energy_balance_interface_tan_comp} as a function of $U_m$ for $Eo=20$, 180, and 300, shown in (a)-(c), respectively, with $Nf = 80$. In each panel, the vertical dashed line marks the $U_m$ value for which $\beta_R=0$. } 
     \label{fig:ls_tanEnergyNf80EoL3M1}
\end{figure}

In Figure  \ref{fig:ls_tanEnergyNf80EoL3M1}, we plot the dependence of the constituents of $TAN$ on $U_m$ for $m=1$, $Eo=20$, 180, and 300, and $Nf=80$. 
Inspection of Figure  \ref{fig:ls_tanEnergyNf80EoL3M1}(a) reveals that in the $Eo=20$ case, for which surface tension effects are important, the major contributor to $TAN$ corresponds to $TAN_{strs}$ and exerts a destabilising influence on the bubble motion.
As can also be seen in Figure \ref{fig:ls_tanEnergyNf80EoL3M1}a, although $TAN_{strs}$ remains destabilising, it gives way to $TAN_g$ as the dominant contributor to $TAN$ with decreasing magnitude of $U_m$, while $TAN_{pb}$ is sufficiently stabilising so as to render $\beta_R <0$; the contributions of $TAN_{ut}$ and $TAN_{ts}$ are relatively negligible and they play an insignificant role in the bubble stability. 

\begin{figure} 
    \centering
    \begin{subfigure}[h]{0.48\linewidth}
        \includegraphics[width=\linewidth]{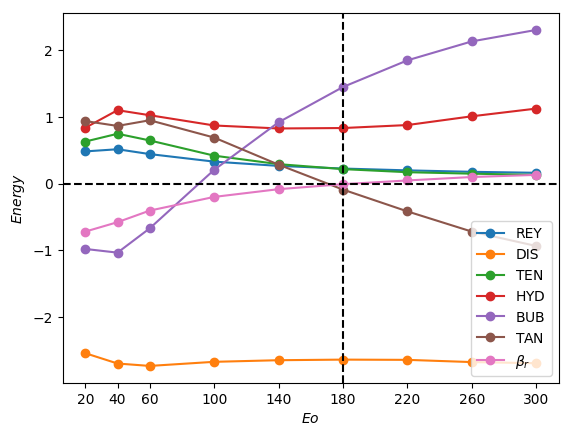}
        \caption{ }
        \label{fig:ls_energyNf80ULmaxdw20L3M1}
    \end{subfigure}
    \begin{subfigure}[h]{0.48\linewidth}
        \includegraphics[width=\linewidth]{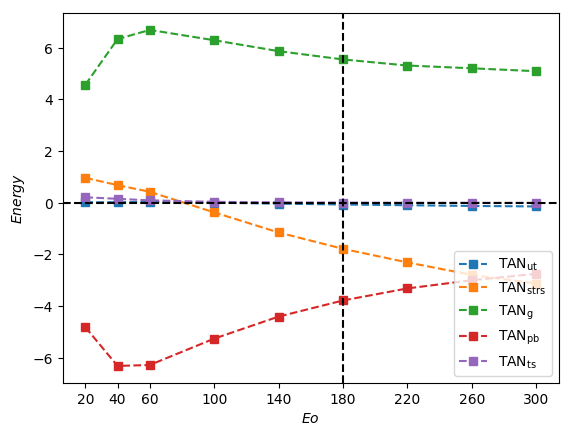}
        \caption{}
         \label{fig:ls_tanEnergyNf80ULmaxdw20L3M1}
    \end{subfigure}
     \caption{Breakdown of the energy budget (see equation \eqref{eq:ls_energy_balance_eqn_scaled}), (a), and the $TAN$ constituents (see equation \eqref{eq:ls_energy_balance_interface_tan_comp}), (b), with $Eo$, for $m=1$ with $Nf = 80$ and $U_m = -0.20$. The vertical dashed line marks the $Eo$ value for which $\beta_R=0$.} 
     \label{fig:ls_tanEnergyNf80ULmaxdw20L3}
\end{figure}

In Figure \ref{fig:ls_tanEnergyNf80EoL3M1}(b,c) generated for $Eo=180$ and 300 for which surface tension effects are weak, the dominant destabilising contribution to $TAN$ is due to $TAN_g$ with the sub-dominant $TAN_{strs}$ and $TAN_{pb}$ exerting a stabilising influence over the majority of the $U_m$ range investigated. The reversal in the role of $TAN_{strs}$ as we cross over from relatively low to high $Eo$ values shown in Figure \ref{fig:ls_tanEnergyNf80EoL3M1}(b,c) is consistent with the results discussed in the previous sections which indicated that viscous effects are destabilising (stabilising) for low (high) $Eo$. This is further illustrated in Figure \ref{fig:ls_tanEnergyNf80ULmaxdw20L3} in which we plot the breakdown of the energy budget (see Figure \ref{fig:ls_tanEnergyNf80ULmaxdw20L3}a) and the constituents of $TAN$ (see Figure \ref{fig:ls_tanEnergyNf80ULmaxdw20L3}b) as a function of $Eo$ for $m=1$ with $U_m=-0.2$ and $Nf=80$. It is clear that $TAN_{strs}$ switches roles in the $Eo$ interval $\left( 60,100 \right]$ and $TAN$ exhibits a similar behaviour over a somewhat larger $Eo$ range.  

\begin{figure} 
    \centering
    \begin{subfigure}[h]{0.48\linewidth}
        \includegraphics[width=\linewidth]{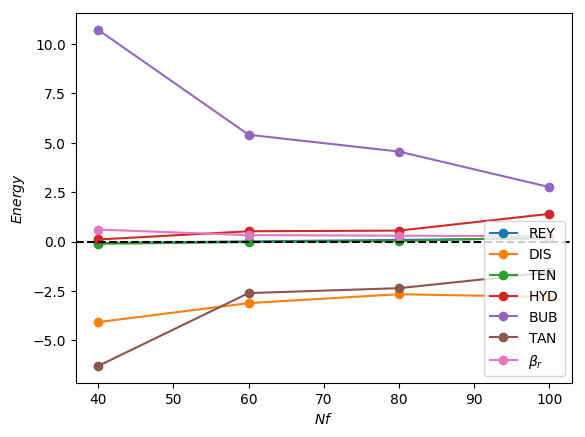}
        \caption{ }
        \label{fig:ls_energyEo300ULmaxdw25L3M1}
    \end{subfigure}
    \begin{subfigure}[h]{0.48\linewidth}
        \includegraphics[width=\linewidth]{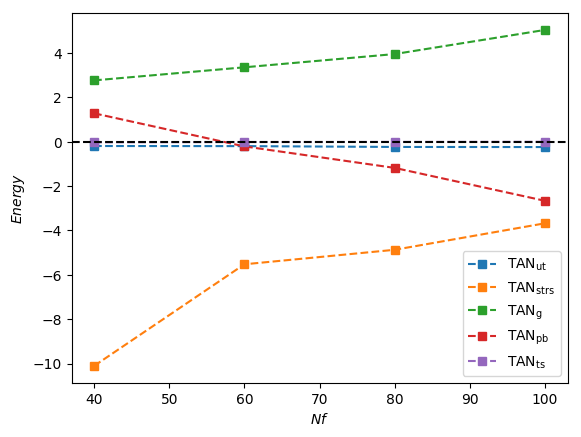}
        \caption{}
         \label{fig:ls_tanEnergyEo300ULmaxdw25L3M1}
    \end{subfigure}
     \caption{Breakdown of the energy budget (see equation \eqref{eq:ls_energy_balance_eqn_scaled}), (a), and the $TAN$ constituents (see equation \eqref{eq:ls_energy_balance_interface_tan_comp}), (b), with $Nf$, for $m=1$ with $Eo=300$ and $U_m=-0.25$. 
     } 
     \label{fig:ls_tanEnergyEo300ULmaxdw25L3}
\end{figure}

Lastly, we show in Figure \ref{fig:ls_tanEnergyEo300ULmaxdw25L3}a breakdown of the energy budget and of the $TAN$ constituents as a function of $Nf$ for $m=1$ with $Eo=300$ and $U_m=-0.25$. It is seen clearly in Figure 14a that for this large $Eo$ case, $BUB$ provides the dominant destabilising contribution with $TAN$ and $DIS$ inducing stability. Inspection of Figure  \ref{fig:ls_tanEnergyEo300ULmaxdw25L3}b reveals that although $TAN_g$ is destabilising over the range of $Nf$ studied, $TAN_{pb}$ is also destabilising for $Nf < 60$; this acts to reduce the stabilising effect associated with the increase in viscous effects and reduction in $Nf$. 

We now establish a connection with the work of \cite{Lu_Prosperetti_2006} who concluded that it is the normal component of gravity on the interface that drives the transition to asymmetric Taylor bubble shape. It is worth mentioning, however, that the analysis of
\cite{Lu_Prosperetti_2006} was carried out locally around the nose region under the assumptions that the effects of viscosity and surface tension are negligible. 
In contrast, our analysis shows that the dominant destabilising mechanisms depend on the relative significance of surface tension characterised by $Eo$:
for low and high $Eo$, the tangential stress, $TAN$ (with $TAN_{strs}$ being the main contributor) and the work done at the interface against the bubble pressure, $BUB$, are chiefly responsible for instability, respectively.  
A look at Figures \ref{fig:ls_energyNf80ULmaxdw20L3M1} and  \ref{fig:ls_energyEo300ULmaxdw25L3M1} shows that the energy term due to normal component of gravity on the interface, $HYD$, is an increasing function of $Nf$ and $Eo$. It is plausible that at very high $Nf$ and $Eo$ $HYD$ may overtake $BUB$ as the most dominant destabilising energy term. It is therefore  likely that the mechanisms governing the instability in regimes of extremely negligible surface tension and viscosity is different from that identified in this investigation. 
%
\section{Summary and conclusions} \label{sec:linear_stability_summary}
We have examined the linear stability of Taylor bubbles in stagnant and flowing liquids in vertical pipes focusing on the case of downward liquid flow. 
The base state, characterised by constant bubble and axisymmetric shapes, was computed by \cite{Abubakar_Matar_2021_s} as a function of 
the E\"{o}tv\"{o}s and inverse viscosity numbers, $Eo$ and $Nf$, and the (centreline) speed of the downward flowing liquid, $U_m$. 
A finite element linear stability model was derived using the concepts of domain perturbation and total linearisation  method presented in \cite{carvalho_scriven_1999} and \cite{Kruyt_etal_1988}, respectively. The model was validated by comparing its predictions  with analytical results for the growth rate and  frequency for the normal mode small-amplitude oscillation of a spherical bubble in an unbounded stagnant liquid under the assumption of negligible gravity.

Our linear stability framework was then used to examine the stability of the base states obtained by \cite{Abubakar_Matar_2021_s}. 
Our results demonstrated that the leading unstable mode corresponds to $m = 1$, where $m$ is the azimuthal wavenumber of the applied perturbation. We constructed stability maps showing the dependence of the critical magnitude of $U_m$ on $Eo$, with $Nf$ varying parametrically, which demarcate the regions in $(U_m,Eo)$ space wherein the flow is linearly unstable.   
At low $Eo$, for which surface tension effects are significant, the instability targets an undulation in the bottom region of the bubble with the three-dimensional bubble shape exhibiting an asymmetric bulge in that region. For weak surface tension effects, characterised by high $Eo$, the most unstable mode corresponds to a deflection of the bubble nose away from the axis of symmetry. 
The stability maps also show the locus of points for which the bubble are stationary in a downward flowing liquid and highlights the regions in parameter space in which they are linearly stable or unstable resulting in axisymmetric and asymmetric shapes, respectively. 

To elucidate the origins of the transition to linear instability and asymmetric bubble shapes, an energy budget analysis was performed to analyse the contribution of various physical mechanisms to the production of perturbation energy.  
This analysis showed that the major contribution to energy production that drives the instability comes from the bubble pressure and the tangential stress for high and low $Eo$ values, respectively. The insights gained from the energy analysis, and the trends observed in the linear stability characteristics, were used to establish clear connections to the influence of $Eo$, $Nf$, and $U_m$ on the curvature of the bubble nose, which plays a crucial role in the stability of the flow.  \\ 

Declaration of Interests. The authors report no conflict of interest. \\

\subsection*{Acknowledgements}

This work was supported by the Engineering $\&$ Physical Sciences Research Council UK through the EPSRC MEMPHIS (grant no. EP/K003976/1) and PREMIERE (grant no. EP/T000414/1) Programme Grants, and a Nigerian government PTDF scholarship for HA. OKM also acknowledges funding from PETRONAS and the Royal Academy of Engineering for a Research Chair in Multiphase Fluid Dynamics, and the PETRONAS Centre for Engineering of Multiphase Systems. We also acknowledge HPC facilities provided by the Research Computing Service (RCS) of Imperial College London for the computing time. 

\appendix
\section{Weak  formulations}\label{sec:weak_formulation}
The transformation of the governing equations into their weak forms involves three steps: multiplying the governing equations for each variables with their corresponding \textit{test functions} and integrating over the domain, integrating by part to reduce the order of integration, and, finally, incorporating the boundary conditions into the resulting relations. Before proceeding to derive the weak forms of the equations, it is important to define the necessary functional spaces to which the solution and test functions must belong \citep{Heinrich_Pepper_1999}: 
\begin{enumerate}
\item The $L^2 \LB \Omega \RB $ space: This is a space of functions $ f \LB \mathbf{r} \RB $ defined in $\Omega$ that are square integrable over $\Omega$: 
\begin{equation}
L^2 \LB \Omega \RB = \left \{ f\LB \mathbf{r} \RB | \int_\Omega \LB f\LB \mathbf{r} \RB \RB ^2 d \Omega < \infty \right \}.
\end{equation}
\item The $L^2_0 \LB \Omega \RB $ subspace: This is a subspace of $L^2 \LB \Omega \RB$ defined in $\Omega$ such that for functions defined in $L^2 \LB \Omega \RB$, the following equation is satisfied:
\begin{equation}
L^2_0 \LB \Omega \RB = \left \{ f | f \in L^2 \LB \Omega \RB  \mbox{ and }  \int_\Omega f\LB \mathbf{r} \RB = \mathrm{0} \right \}.
\end{equation}
\item The Sobolev space $H^1  \LB \Omega \RB$: this is a space of functions $ f \LB \mathbf{r} \RB $ defined in $\Omega$ such that both the function and all its first partial derivatives are in $L^2 \LB \Omega \RB $ 
\begin{equation}
H^1 \LB \Omega \RB = \left \{ f\LB \mathbf{r} \RB | \int_\Omega \LS   |f|^2 + |\nabla f| ^2 \RS  d \Omega < \infty \right \}.
\end{equation}
\item The Sobolev subspace $H^1_0  \LB \Omega \RB $: this is a subspace of the Sobolev space $H^1  \LB \Omega \RB$  for which the functions defined in space $H^1$ vanish on the portions of the boundary of  $\Omega$ where \textit{Dirichilet} boundary conditions are imposed ( i.e $\Gamma_D = \Gamma_{in} + \Gamma_{wall} $ ): 
\begin{equation}
H^1_0 \LB \Omega \RB = \left \{ f | f \in H^1 \LB \Omega \RB  \mbox{ and }  f\LB \mathbf{r} \RB = \mathbf{0} \mbox{ if } \mathbf{r} \in \Gamma_D \right \}.
\end{equation}
\end{enumerate}
Let $\mathbf{\Phi} \in H^1_0 $ and $ \varphi \in L^2_0 $ be the {test functions}  corresponding to $\mathbf{u} \in H^1 $ and $\mathrm{p} \in L^2$, respectively.
Next we take the inner product of equation (\ref{eq:momentum}) with $\mathbf{\Phi} $ , multiply equation (\ref{eq:continuity}) with $ \varphi $ and integrate the equations over the domain:
%
%
%
\begin{equation}
\int_{\Omega}
\left\{
\frac{\partial {\mathbf{u}}}{\partial t} \cdot \mathbf{\Phi} 
+
\left[ 
\left({\mathbf{u}} \cdot \nabla \right)\mathbf{u}
\right] 
\cdot \mathbf{\Phi}
-
\LS \nabla \cdot \mathbf{T} \RS \cdot \mathbf{\Phi}
\right\}d\Omega
=0,  \label{eq:weak_form_momentum0}
\end{equation}
\begin{equation}
\int_{\Omega}
\left\{
 \left( \nabla \ldotp \mathbf{u} \right)\varphi 
 \right\}d\Omega = 0.  \label{eq:weak_form_continuity} 
 \end{equation}
Equations (\ref{eq:weak_form_momentum0}) and (\ref{eq:weak_form_continuity}) are the weighted residual forms of the momentum and continuity equations, respectively. Integrating the last term on the left-hand-side of (\ref{eq:weak_form_momentum0}) by parts, 
\begin{equation}
\int_{\Omega}
\left\{
\LS \nabla \cdot \mathbf{T} \RS \cdot \mathbf{\Phi} 
\right\}d\Omega
= 
\int_{\Gamma }
\left\{ 
\mathbf{n}\cdot\mathbf{T}\cdot \mathbf{\Phi}
\right\}d\Gamma_b
-
\int_{\Omega}
\left\{
\mathbf{T} : \nabla \LB \mathbf{u}\RB
\right\}d\Omega,
\end{equation}
where
\begin{equation*}
\Gamma = \Gamma_b + \Gamma_{in} + \Gamma_{wall} + \Gamma_{out}.
\end{equation*} 
Enforcing the outlet boundary condition  and taking into consideration that $\mathbf{\Phi} \in H^1_0$, hence $\mathbf{\Phi}$ are zero where \textit{essential} boundary conditions are imposed, we are left with 
\begin{equation*}
\Gamma = \Gamma_b.
\end{equation*}
Making use of (\ref{eq:stress_tensor}), equation (\ref{eq:weak_form_momentum0}) becomes
\begin{multline}
\int_{\Omega}
\left\{
\frac{\partial {\mathbf{u}}}{\partial t} \cdot \mathbf{\Phi} 
+
\left[ 
\left({\mathbf{u}} \cdot \nabla \right)\mathbf{u}
\right] 
\cdot \mathbf{\Phi}
+
2 {Nf}^{-1} \mathbf{E}\left({\mathbf{u}} \right) 
: \mathbf{E}\left(\mathbf{\Phi} \right)
-
{p} \left( \nabla \ldotp \mathbf{\Phi} \right)
\right\}d\Omega 
\\
-
\int_{\Gamma_b}
\left\{ 
\mathbf{n}\cdot\mathbf{T}\cdot \mathbf{\Phi}
\right\}d\Gamma_b
=0.    \label{eq:weak_form_momentum1}
\end{multline}
The traction term in the last term on the left-hand-side of (\ref{eq:weak_form_momentum1}) can be decomposed into its normal and tangential components \citep{Pozrikidis_2011}:
\begin{equation}
\mathbf{n} \cdot {\mathbf{T}} = \left[\mathbf{n}\cdot {\mathbf{T}}\cdot \mathbf{n} \right] \mathbf{n} + \mathbf{n} \times \left[ \mathbf{n}\cdot {\mathbf{T}} \times \mathbf{n} \right], \label{eq:cd_traction_decomposition}
\end{equation}
thereby allowing the incorporation of normal stress condition (\ref{eq:normal_stress_bc}) and tangential stress condition (\ref{eq:tangential_stress_bc}) into (\ref{eq:weak_form_momentum1}) to give
\begin{align}
%
& \int_{\Omega}
\left\{
\frac{\partial {\mathbf{u}}}{\partial t} \cdot \mathbf{\Phi} 
+
\left[ 
\left({\mathbf{u}} \cdot \nabla \right)\mathbf{u}
\right] 
\cdot \mathbf{\Phi}
+
2 {Nf}^{-1} \mathbf{E}\left({\mathbf{u}} \right) 
: \mathbf{E}\left(\mathbf{\Phi} \right)
-
{p} \left( \nabla \ldotp \mathbf{\Phi} \right)
\right\}d\Omega \nonumber \\
& \quad {} -
\int_{\Gamma_b}
\left\{ 
\left[ 
{Eo}^{-1}\kappa  + \mathrm{z}  - \mathrm{P}_{\mathrm{b}}
\right] \mathbf{n} \cdot \mathbf{\Phi}
\right\}d\Gamma_b
=0.  \label{eq:weak_form_momentum2}
\end{align}
Equations   \eqref{eq:weak_form_continuity} and \eqref{eq:weak_form_momentum2} are the weak forms of the governing equations.
\section{Curvature linearisation} \label{sec:curvature_linearisation}
For a three-dimensional axisymmetric surface, the interface location in any $(r,z)$ plane is sufficient for computing the curvature of the surface. Consider that the interface in any such plane is spanned by a curve with the coordinates of any point on the curve being $(r,z)$. In addition, let the interface be parametrised by length of arc $s$ so that the position vector of any point on the interface is given as 
\begin{equation}
\mathbf{r}= r(s)\mathbf{i}_r + z(s) \mathbf{i}_z.
\end{equation}
The total curvature at any given point on the interface is defined as 
\begin{equation}
\kappa = -\nabla_s \cdot \mathbf{n}, 	 \label{eq:curvature_formular}
\end{equation}   
where $r \LB s\RB $ and $z \LB s\RB $ are the radial and axial coordinates of the points on the interface, respectively; $\mathbf{n}$  remains the unit normal to the interface and $\nabla_s$ is the surface tangential gradient operator which is given as
\begin{equation}
\nabla_s = \LB \mathbf{I} - \mathbf{n} \otimes \mathbf{n}\RB \nabla.	 \label{eq:surface_div}
\end{equation}
For a three-dimensional axisymmetric surface, (\ref{eq:surface_div}) simplifies to 
\begin{equation}
\nabla_s = t_r \frac{d}{ds} \mathbf{i}_r + \frac{1}{r} \frac{\partial}{\partial \theta}\mathbf{i}_\theta +  t_z \frac{d}{ds} \mathbf{i}_z, \label{eq:surface_div1}
\end{equation}
and equation \eqref{eq:curvature_formular}  becomes
\begin{equation}
\kappa = -\LB \mathbf{t} \cdot \frac{d \mathbf{n}}{d s} + \frac{n_r}{r} \RB,  \label{eq:curvature_expanded_form}
\end{equation}
where $\mathbf{t}$ is the unit tangent vector to the interface with components $t_r$, $t_\theta$ and $t_z$ in the radial, azimuthal and axial directions, respectively; $\frac{d}{ds} = \left( \mathbf{t} \cdot \nabla \right)$ is an operator that denotes the derivative in the tangential direction; $n_r$ is the radial component of the unit normal vector, $\mathbf{n}$. 

Let us imagine that the  deformed interface can be expressed as a summation of the undeformed interface and a very small deformation. Thus, the deformed interface can be written as 
\begin{equation}
\mathbf{r}  = \mathbf{r}^0 + {\mathbf{x}},		\label{eq:interface_deform}
\end{equation} 
where $\mathbf{r}^0 = r^0 \mathbf{i}_r + \theta^0 \mathbf{i}_\theta + z^0 \mathbf{i}_z $  and  $\mathbf{r} = r \mathbf{i}_r + \theta \mathbf{i}_\theta + z \mathbf{i}_z $  are the undeformed (i.e base) and deformed (i.e perturbed) interface position vectors, respectively; $\mathbf{x} = x_r \mathbf{i}_r + x_\theta \mathbf{i}_\theta + x_z \mathbf{i}_z $, as mentioned in the previous section, is the interface deformation vector and  is taken to be very small in magnitude. 
Linearisation of the unit normal to,  and elemental arc length on, the deformed interface about the undeformed interface give \citep{Ramanan_Engelman_1996,Kruyt_etal_1988,Weatherburn_1927}
\begin{equation}
\mathbf{n} = \mathbf{n}^0 -  \mathbf{n}^0 \times \nabla_s  \times \mathbf{x},		\label{eq:normal_linearised}
\end{equation}
\begin{equation}
ds = \LB 1 +  \mathbf{t}^0 \cdot \frac{d \mathbf{x}}{ds^0}\RB ds^0, \label{eq:arc_length_linearised}
\end{equation}
\begin{equation}
\frac{d }{ds}  = \LB 1 -  \mathbf{t}^0 \cdot \frac{d \mathbf{x}}{ds^0}\RB  \frac{d }{ds^0}. \label{eq:arc_length_deriva_linearised}
\end{equation}
On  further simplification of \eqref{eq:normal_linearised}  
\begin{equation}
\mathbf{n} = \mathbf{n}^0 - \mathbf{t}^0 \LB \mathbf{n}^0 \cdot \frac{d \mathbf{x}}{d s^0}\RB - \frac{\mathbf{n}^0}{r^0} \cdot \frac{\partial \mathbf{x}}{\partial \theta^0} \mathbf{i}_\theta.  \label{eq:normal_linearised_simplified_a}
\end{equation}
Substituting  \eqref{eq:normal_linearised_simplified} together with \eqref{eq:interface_deform} and \eqref{eq:arc_length_deriva_linearised} into \eqref{eq:curvature_expanded_form}, the linearised curvature neglecting all terms of nonlinear in $x$ gives
\begin{equation}
\kappa = \kappa^0 +  \kappa^1,	\label{eq:curvature_split}
\end{equation}
with
\begin{equation}
\kappa^0 =  -\LB \mathbf{t}^0 \cdot \frac{d \mathbf{n}^0}{d s^0} + \frac{n_r^0}{r^0} \RB,  \label{eq:curvature_base}
\end{equation}
\begin{multline}
\kappa^1 = \frac{1}{r^0} \frac{d}{d s^0} \LS r^0 \LB \mathbf{n}^0 \cdot \frac{d \mathbf{x}}{d s^0}\RB \RS + 2 \LB \mathbf{t}^0 \cdot  \frac{d \mathbf{x}}{d s^0}  \RB \LB \mathbf{t}^0 \cdot \frac{d \mathbf{n}^0}{d s^0} \RB + \frac{\mathbf{n}^0}{{r^0}^2} \cdot \frac{\partial^2 \mathbf{x}}{\partial {\theta^0}^2} + \frac{x_r n_r^0}{{r^0}^2}
\\
 - \frac{d \mathbf{n}^0}{d s^0} \cdot \frac{d \mathbf{x}}{d s^0}.  \label{eq:curvature_perturb0_a}
\end{multline}
When \eqref{eq:curvature_perturb0} is further simplified by allowing the deformation vector to be of the form 
\begin{equation}
\mathbf{x} = h \mathbf{n}^0, \label{eq:deformation_form}
\end{equation}
it results in
\begin{equation}
\kappa^1 =  \frac{1}{r^0} \frac{d}{d s^0} \LB r^0 \frac{d h}{d s^0} \RB + h \LS {\kappa^0}^2_a + {\kappa^0}^2_b +  \frac{1}{{r^0}^2} \frac{\partial^2 h}{\partial {\theta^0}^2} \RS, \label{eq:curvature_perturb1}
\end{equation}
which is the same as the expression derived for curvature perturbation in \cite{Chireux_etal_2015}, albeit through a different and longer route.
In arriving at \eqref{eq:curvature_perturb1}, we have used the following Frenet-Serret relations 
\begin{subequations} \label{eq:frenet_serret}
\begin{align}
\frac{d \mathbf{t} }{d s} &= \kappa \mathbf{n}, \\
\frac{d \mathbf{n} }{d s} &= - \kappa \mathbf{t}.
\end{align}
\end{subequations}
Equations \eqref{eq:curvature_perturb0_a} and \eqref{eq:curvature_perturb1} are the expressions for curvature deformation and can be used for an axisymmetric deformation by setting the term containing derivative with respect to the azimuthal coordinate to zero to obtain
\begin{equation}
\kappa^1 =  \frac{1}{r^0} \frac{d}{d s^0} \LB r^0 \frac{d h}{d s^0} \RB + h \LS {\kappa^0}^2_a + {\kappa^0}^2_b \RS. \label{eq:curvature_perturb_axis_2d}
\end{equation}
In equations \eqref{eq:normal_linearised}-\eqref{eq:curvature_perturb1}, $\mathbf{n}^0$ is the unit normal vector to the undeformed interface and $n_r^0 $, $n_\theta^0$ and $n_z^0 $ are its component in the radial, azimuthal and axial directions, respectively; $\mathbf{t}^0$ is the unit tangent vector to the undeformed interface with components  $t_r^0 $, $t_\theta^0 $ and  $t_z^0 $ in the radial, azimuthal and axial directions, respectively; $ds$ and $ds^0$ are the elemental arc length for the deformed and undeformed interfaces, respectively; $\kappa^0$ is the curvature of the undeformed surface and $\kappa^1$ is the addition to the undeformed interface curvature (also referred to as curvature perturbation in the context of linear stability analysis) due to linearisation of the deformed interface about the undeformed interface; $h$ is the magnitude of the interface deformation in the direction normal to the undeformed interface. ${\kappa^0}_a$ and ${\kappa^0}_b$ are the two principal curvatures of the undeformed interface \eqref{eq:curvature_base} corresponding to the curvature in the $r-z$  and $r-\theta$ planes, respectively, defined as
\begin{subequations}
\begin{align}
{\kappa^0}_a &= \mathbf{t}^0 \cdot \frac{d \mathbf{n}^0}{d s^0}, \label{eq:curvature_k0a} \\
 {\kappa^0}_b &= \frac{n_r^0}{r^0}.	\label{eq:curvature_k0b}
\end{align}
\end{subequations}

\section{Galerkin finite element approximation} \label{sec:galerkin_method}
In appendix \ref{sec:weak_formulation}, we derived the weak forms of the governing equations in the continuous domain. Let us divide the domain, $\Omega$ into smaller subsets, \textit{finite element}, so that each element occupies a sub-domain $\Omega^e$ with boundary $\Gamma^e$.
The discretised domain is therefore an assemblage of $n_e$ finite elements that make up the domain:
\begin{equation}
\int_{\Omega^h} \left \{ \cdot \right \} d \Omega^h = \sum_{k = 1}^{ne} \int_{\Omega^e_k} \left \{ \cdot \right \} d \Omega^e_k,
\end{equation}
\begin{equation}
\int_{\Gamma^h} \left \{ \cdot \right \} d \Gamma^h = \sum_{k = 1}^{ne} \int_{\Gamma^e_k} \left \{ \cdot \right \} d \Gamma^e_k,
\end{equation}
where $\Omega^h$ represents the discretised domain with boundary $\Gamma^h$, and $ne$ is the number of elements in the discretised domain.
\begin{figure}
\centering
	\includegraphics[width = 0.65\linewidth]{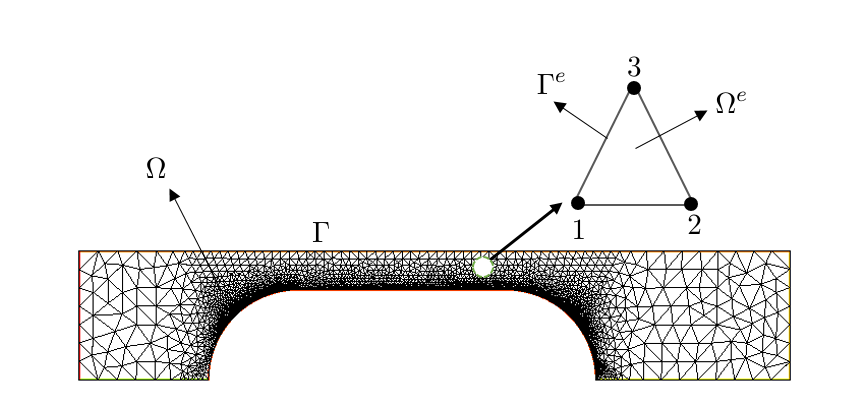}
	\caption{Two-dimensional triangulated domain with linear approximations on the subdomains thereby leading to three nodes.} 
	\label{fig:domain_discretised}
\end{figure}
The geometry of the finite element could be a triangle or quadrilateral in a two-dimensional domain and in three-dimensional domain, it could be a tetrahedral or hexahedral.
On each element, the unknown variables of the problem, known as \textit{trial functions}, are approximated as a linear combination of unknown parameters and piecewise polynomial functions,  \textit{basis}  (or \textit{shape} ) \textit{functions}. In addition, the test functions corresponding to each variables are set to equal to the basis functions used in approximating the variables. Depending on the geometry of an element and the order of polynomial used in approximating the unknown variable on the element,  $n_v$ number of nodes/unknown parameters can be associated with the element as shown in Figure \ref{fig:domain_discretised}. 
Thus, on each element, the trial  and test functions are approximated as
\begin{subequations} \label{eq:trial_functions_aprox}
\begin{align}
\mathrm{ u_r \LB \mathbf{x}, t \RB } &=  \sum_{k = 1}^{n_v} \psi^k\LB \mathbf{x}\RB  v_r^k \LB  t \RB = \mathbf{\Psi}^T \mathbf{v}_r,\\
 \mathrm{u}_\theta \LB \mathbf{x}, t \RB  &=  \sum_{k = 1}^{n_v} \psi^k\LB \mathbf{x}\RB  v_{\theta}^k \LB  t \RB = \mathbf{\Psi}^T \mathbf{v}_\theta, \\
 \mathrm{u}_z \LB \mathbf{x}, t \RB  &=  \sum_{k = 1}^{n_v} \psi^k\LB \mathbf{x}\RB  v_{z}^k \LB  t \RB = \mathbf{\Psi}^T \mathbf{v}_z,\\
 \mathrm{p} \LB \mathbf{x}, t \RB  &=  \sum_{l = 1}^{n_p} \lambda^l\LB \mathbf{x}\RB  \mathfrak{p}^l \LB  t \RB  = \mathbf{\Lambda}^T \mathbf{p}, \\
 h &= \sum_{k = 1}^{n_h} \eta_k \LB \mathbf{r}_b \RB  \hslash_k \LB  t \RB = \text{\boldmath ${\eta}$}^T \mathbf{h},  \label{eq:trial_functions_aprox_press}
\end{align}  
\end{subequations}
\begin{subequations}
\begin{align}
 \label{eq:test_functions_aprox}
\Phi_r = \psi^k, \quad \Phi_\theta = \psi^k, \quad \Phi_z = \psi^k \quad  \forall \quad k = 1, \cdots, n_v;  \\
\varphi = \lambda^l \quad \forall \quad l = 1, \cdots, n_p; \\ \xi = \eta_m \quad  \forall \quad m = 1, \cdots, n_h. 
\end{align}
\end{subequations}
In equations  (\ref{eq:trial_functions_aprox}) and (\ref{eq:test_functions_aprox}), $\psi$, $\lambda$  and $\eta$ are the shape functions for the velocity components, pressure, and interface deformation magnitude, respectively; $n_v$, $n_p$ and $n_h$ are the number of nodes for the velocity components, pressure, and interface deformation magnitude, respectively, and $v_j \; \forall \; j = r,\; \theta, \; z $ , $\mathfrak{p}$ and $\hslash$ are the corresponding unknown nodal parameters of the velocity components, pressure, and interface deformation magnitude. $\mathbf{v}_i \;  \forall \; i = r,\; \theta \; z $,  $\mathbf{p} $  and $\mathbf{h}$ are column vectors of the unknown nodal parameters for the velocity components, pressure, and interface deformation magnitude, respectively; and  $\mathbf{\Psi}$, $\mathbf{\Lambda}$ and $\text{\boldmath ${\eta}$}$ are the column vectors of the shape functions for the velocity components, pressure, and interface deformation magnitude, respectively. 

It is seen that the shape functions for the velocity components are the same but differ from that of the pressure. This is to avoid having an over-constrained system of discrete equations. In fact, the shape function used for pressure must be at least one order lower than that of the velocity field \citep{Reddy_Gartling_2010}. 

Using equations (\ref{eq:trial_functions_aprox}) and (\ref{eq:test_functions_aprox}), we can derive, for example, the finite element model for the continuity equation by writing the  weak form of it \eqref{eq:ls_weak_form_normal_mode_perturbation_continuity} over an element and substituting (\ref{eq:trial_functions_aprox}) and (\ref{eq:test_functions_aprox}) to get
\begin{multline}
\LS
\int_{\Omega^e}
\left\{
\mathbf{\Lambda}
\LB
\frac{\partial \mathbf{\Psi}^T}{\partial r} + \frac{\mathbf{\Psi}^T}{r} 
\RB
\right\}d\Omega^e 
\RS \mathbf{v}_r
+ 
\LS
\int_{\Omega^e}
\left\{
\mathfrak{i} m {\mathbf{\Lambda}} \frac{{\mathbf{\Psi}}^T}{r}
\right\}d\Omega^e
\RS \mathbf{v}_\theta
\\
+ 
\LS
\int_{\Omega^e}
\left\{
\mathbf{\Lambda} \frac{\partial \mathbf{\Psi}^T}{\partial z}
\right\}d\Omega^e
\RS \mathbf{v}_z
= \mathbf{0},
\end{multline}
which can be compactly written as
\begin{equation}
\mathbf{Q}^T \mathbf{v} = \mathbf{0}, \label{eq:gfem_continuity}
\end{equation} 
where $\mathbf{v} = 
\begin{bmatrix}
\mathbf{v}_r \\
\mathbf{v}_\theta \\
\mathbf{v}_z 
\end{bmatrix} $ and $\mathbf{Q} = 
\begin{bmatrix}
\mathbf{Q}_r \\
\mathbf{Q}_\theta \\
\mathbf{Q}_z 
\end{bmatrix} $
. $\mathbf{Q}_i \quad \forall \; i = r,\; \theta \; z  $ are the coefficient matrices for the velocity components.
Similar compact relations to \eqref{eq:gfem_continuity} can be derived  for the momentum equations and the kinematic boundary condition. The combined formed of which is given in equation  \eqref{eq:ls_algebraic_sys_eqns0} and full expressions for the terms are: \\
\textit{continuity equation}
\begin{align}
& {\mathbf{Q}}_r 
 = 
 \int_{\Omega^{0,e}}
\left\{
\mathbf{\Lambda}
\LB
\frac{\partial {\mathbf{\Psi}}^T}{\partial r} + \frac{{\mathbf{\Psi}}^T}{r} 
\RB
\right\}d\Omega^{0,e},
\\
& {\mathbf{Q}}_\theta 
= 
\int_{\Omega^{0,e}}
\left\{
\mathfrak{i} m {\mathbf{\Lambda}} \frac{{\mathbf{\Psi}}^T}{r}
\right\}d\Omega^{0,e}, \\
& {\mathbf{Q}}_z 
= 
\int_{\Omega^{0,e}}
\left\{
\mathbf{\Lambda} \frac{\partial {\mathbf{\Psi}}^T}{\partial z}
\right\}d\Omega^{0,e}, \\
&\mathbf{Q}_p 
 = 
 \int_{\Omega^{0,e}}
\left\{
\varepsilon
\mathbf{\Lambda}
\mathbf{\Lambda}^T
\right\}d\Omega^{0,e}; 
\end{align}
\textit{momentum equation: growth rate part}
\begin{align}
{\mathbf{B}}_{i,i} 
&=
\int_{\Omega^{0,e}}
\left\{
{\mathbf{\Psi}} {\mathbf{\Psi}}^T  
\right\} d\Omega^{0,e}  \quad \forall \; i = r,\theta, z,
\end{align}
\textit{momentum equation: convective part}
\begin{align}
\mathbf{C}_{i,i} 
 &= 
 \int_{\Omega^{0,e}}
\left\{
\mathbf{\Psi}
\left[ 
\left({\mathbf{u}^0} \cdot \nabla \right)\mathbf{\Psi}^T 
+ \delta_{r,i} \mathbf{\Psi}^T \frac{\partial u_r^0}{\partial r}
+ \delta_{z,i} \mathbf{\Psi}^T \frac{\partial u_z^0}{\partial z}
\right] 
\right\}d\Omega^{0,e}  \; \forall \; i = r,z, \\
{\mathbf{C}}_{\theta,\theta} 
 &= 
 \int_{\Omega^{0,e}}
\left\{
{\mathbf{\Psi}}
\left[ 
\left({\mathbf{u}^0} \cdot \nabla \right) {\mathbf{\Psi}}^T 
+ \frac{ u_r^0 {\mathbf{\Psi}}^T}{r} 
\right] 
\right\}d\Omega^{0,e},  \\
\mathbf{C}_{r,z} 
 &= 
\int_{\Omega^{0,e}}
\left\{
\mathbf{\Psi}
\mathbf{\Psi}^T \frac{\partial u_r^0}{\partial z}
\right\}d\Omega^{0,e},   \\
\mathbf{C}_{z,r} 
 &= 
\int_{\Omega^{0,e}}
\left\{
\mathbf{\Psi}
\mathbf{\Psi}^T \frac{\partial u_z^0}{\partial r} 
\right\}d\Omega^{0,e};
\end{align}
\textit{momentum equation: viscous part}
\begin{subequations}	
\begin{align}
{\mathbf{K}}_{i,i} 
 &= 
 \int_{\Omega^{0,e}}
 Nf^{-1}
\left\{
2 \frac{\partial {\mathbf{\Psi}}}{\partial i} \frac{\partial {\mathbf{\Psi}}^T}{\partial i} 
+ \LB 2 \delta_{r,i} + m^2 \RB \frac{ {\mathbf{\Psi}} {\mathbf{\Psi}}^T}{r^2}
+ \delta_{r,i} \frac{\partial {\mathbf{\Psi}}}{\partial z} \frac{\partial {\mathbf{\Psi}}^T}{\partial z}
+ \delta_{z,i} \frac{\partial {\mathbf{\Psi}}}{\partial r} \frac{\partial {\mathbf{\Psi}}^T}{\partial r}
\right.  \nonumber \\ 
 & \qquad {} + 
\left.
\delta_{z,i} \frac{\partial {\mathbf{\Psi}}}{\partial r} \frac{\partial {\mathbf{\Psi}}^T}{\partial r}
\right\}d\Omega^{0}  \; \forall \; i = r,z, \\
{\mathbf{K}}_{\theta,\theta} 
 &= 
 \int_{\Omega^{0,e}}
 Nf^{-1}
\left\{
\LB 1 + 2m^2 \RB \frac{ {\mathbf{\Psi}} {\mathbf{\Psi}}^T}{r^2}
+ \frac{\partial {\mathbf{\Psi}}}{\partial z} \frac{\partial {\mathbf{\Psi}}^T}{\partial z}
+  \frac{\partial {\mathbf{\Psi}}}{\partial r} \frac{\partial {\mathbf{\Psi}}^T}{\partial r}
\right.  \nonumber \\ 
 & \qquad {} - 
\left. 
\LB 
\frac{{\mathbf{\Psi}}^T}{r} \frac{\partial {\mathbf{\Psi}}}{\partial r} + \frac{ {\mathbf{\Psi}}}{r} \frac{\partial {\mathbf{\Psi}}^T}{\partial r} \RB
\right\} d\Omega^{0,e}, \\
\mathbf{K}_{r,z} 
 &= 
 \int_{\Omega^{0,e}}
 Nf^{-1}
\left\{
\frac{\partial \mathbf{\Psi}}{\partial z} \frac{\partial \mathbf{\Psi}^T}{\partial r}
\right\}d\Omega^{0,e},  \\
{\mathbf{K}}_{i,\theta} 
 &= 
 \int_{\Omega^{0,e}}
 Nf^{-1}
\left\{
\delta_{r,i} 3 \mathfrak{i} m  \frac{{\mathbf{\Psi}} {\mathbf{\Psi}}^T }{r^2} -  \mathfrak{i} m \frac{{\mathbf{\Psi}}}{ r} \frac{\partial {\mathbf{\Psi}}^T}{\partial i} 
\right\}d\Omega^{0,e}  \; \forall \; i = r,z; 
\end{align}
\end{subequations}
\textit{momentum equation: interface deformation part}
\begin{subequations}
\begin{align}
{\mathbf{M}}_{i,h} 
=&
-\int_{\Gamma_b^{0,e}} 
{Eo}^{-1} 
\left \{
- \left[ n_i  \frac{d {\mathbf{\Psi}} }{ds} - \kappa_a \left( t_i  {\mathbf{\Psi}} \right)\right] \frac{d \text{\boldmath ${{\eta}}$}^T}{ds}
+
 n_i   \left[\kappa_a^2 + \kappa_b^2 - \frac{m^2}{r^2}\right] {\mathbf{\Psi}} \text{\boldmath ${{\eta}}$}^T
\right \} 
d\Gamma_b^{0,e}  \nonumber \\
&-
\int_{\Gamma_b^0}
\left\{ 
n_z n_i  {\mathbf{\Psi}} \text{\boldmath ${\eta}$}^T
\right\}d\Gamma_b^0  \nonumber \\
&+ 
\int_{\Gamma^0_b}
\left\{
\left\{
{
\LB
{u}^0_i {t}_i
\RB
\left[ 
\frac{d {u}^0_i}{d s}  {\mathbf{\Psi}}
\right] 
}
+
\LS -p^0 + 2 Nf^{-1} \LB {t}_i \frac{d {u}^0_i}{d s} \RB  \RS \LB {t}_i  \frac{d {\mathbf{\Psi}}}{d s} \RB 
\right\} \text{\boldmath ${\eta}$}^T
\right\}
d\Gamma^0_b  \nonumber \\
&+
\int_{\Gamma_b^0}
\left\{ 
\left[ 
{Eo}^{-1} \kappa  + z - P_b^0
\right] 
\LS 
\LB {t}_i  {\mathbf{\Psi}} \RB \frac{d \text{\boldmath ${\eta}$}^T }{ds} +
\kappa \LB {n}_i  {\mathbf{\Psi}} \text{\boldmath ${\eta}$}^T \RB
\RS
\right\}d\Gamma_b^0 \quad \forall \; i = r,z, \\
{\mathbf{M}}_{\theta,h} 
=& \int_{\Gamma^0_b}
\left\{
\LS \LB -p^0 + 2 Nf^{-1} \frac{u_r^0}{r} \RB  -
\left( 
{Eo}^{-1} \kappa  + z - P_b^0
\right)  \RS \LB -\mathfrak{i} m  \frac{{\mathbf{\Psi}} \text{\boldmath ${\eta}$}^T}{r} \RB 
\right\}
d\Gamma^0_b;
\end{align}
\end{subequations}
\textit{kinematic boundary condition}
\begin{subequations}
\begin{align}
{\mathbf{X}}_i 
&= 
\int_{\Gamma_b^{0,e}}
n_i 
\left\{
\text{\boldmath ${\eta}$}  \mathbf{\Psi}^T
\right\}
d\Gamma_b^{0,e}
\quad
\; \forall \; i = r,z, \\
{\mathbf{X}}_h 
&= 
\int_{\Gamma_b^{0,e}}
\left\{
\LB \mathbf{u}^0 \cdot \mathbf{t} \RB   \text{\boldmath ${\eta}$} \frac{d \text{\boldmath ${\eta}$}^T}{ds} 
- 
\LB \frac{d \mathbf{u}^0}{dn} \cdot \mathbf{n} \RB \text{\boldmath ${\eta}$} \text{\boldmath ${\eta}$}^T 
\right\}
d\Gamma_b^{0,e}, \\
{\mathbf{B}}_h 
&= 
\int_{\Gamma_b^{0,e}}
\left\{
\text{\boldmath ${\eta}$}  \text{\boldmath ${\eta}$}^T
\right\}
d\Gamma_b^{0,e}.
\end{align}
\end{subequations} 

\bibliographystyle{jfm}

\end{document}